\documentclass[preprint, aoas]{imsart}

\usepackage{lipsum}
\usepackage{amsthm,amsmath,xparse}
\usepackage{xcolor}
\usepackage{epsfig}
\RequirePackage{natbib}
\usepackage{multirow}
\usepackage{subcaption}
\RequirePackage[OT1]{fontenc}
\usepackage{slashbox}
\usepackage{multirow}
\usepackage{hhline}
\RequirePackage[hidelinks,citecolor=black,urlcolor=black]{hyperref}
\usepackage{caption}

\usepackage{algorithmic}
\usepackage[ruled]{algorithm2e}
\usepackage{makecell}
\newcolumntype{H}{>{\setbox0=\hbox\bgroup}c<{\egroup}@{}}
\definecolor{shadecolor}{rgb}{0.5,0.5,0.5}
\definecolor{darkblue}{RGB}{0,0,200}
\usepackage{bm}
\usepackage{bbold}
\usepackage{tikz}
\usepackage{ctable}
\usepackage{graphicx}
\usetikzlibrary{fit,positioning}
\definecolor{grey}{gray}{0.9}
\usepackage{blkarray}
\newtheorem{mydef}{Definition}

\newcommand{\iidsim}{\overset{i.i.d}{\sim}}
\newcommand{\precc}{\!\prec\!}

\begin{document}
\begin{frontmatter}

\title{A Bayesian Mallows approach to non-transitive pair comparison data: \\how human are sounds?} 
\runtitle{A Bayesian Mallows approach to non-transitive data}

\begin{aug}
\author{\fnms{Marta} \snm{Crispino$^{1}$}\ead[label=e1]{marta.crispino@inria.fr}}
\address{\printead{e1}},
\author{\fnms{Elja} \snm{Arjas$^{2,3}$}\ead[label=e2]{elja.arjas@helsinki.fi}}
\address{\printead{e2}},
\author{\fnms{Valeria} \snm{Vitelli$^{3}$}\ead[label=e3]{valeria.vitelli@medisin.uio.no}}
\address{\printead{e3}},
\author{\fnms{Natasha} \snm{Barrett$^{4}$}\ead[label=e5]{nlb@natashabarrett.org}}
\address{\printead{e5}}
\and\author{\fnms{Arnoldo} \snm{Frigessi$^{3,5}$}\ead[label=e4]{frigessi@medisin.uio.no}}
\address{\printead{e4}}
\affiliation{Inria Grenoble$^{1}$\footnote{The first author was a PhD student at Bocconi University, Milan, Italy, and visiting OCBE, University of Oslo, during the project.}, University of Helsinki$^{2}$, University of Oslo$^{3}$,  Norwegian State Academy for Music in Oslo$^{4}$, Oslo University Hospital$^{\, 5}$}
\runauthor{M. Crispino et al.}
\end{aug}

\begin{abstract}
We are interested in learning how listeners perceive sounds as having human origins. An experiment was performed with a series of electronically synthesized sounds, and listeners were asked to compare them in pairs. 
We propose a Bayesian probabilistic method to learn individual preferences from non-transitive pairwise comparison data, as happens when one (or more) individual preferences in the data contradicts what is implied by the others. We build a Bayesian Mallows model in order to handle non-transitive data, with a latent layer of uncertainty which captures the generation of preference misreporting. We then develop a mixture  extension of the Mallows model, able to learn individual preferences in a heterogeneous population. 
The results of our analysis of the musicology experiment are of interest to electroacoustic composers and sound designers, and to the audio industry in general, whose aim is to understand how computer generated sounds can be produced in order to sound more human.

\end{abstract}


\end{frontmatter}

\section{Introduction}\label{intro}

We consider experiments involving a set of assessors (experts, judges, users) who express preferences about a set of items.
Each assessor is shown a predetermined sequence of pairs of items, one pair at a time, and chooses from every pair the item that she prefers. Preference is here interpreted in a broad sense as an order relation. 
The assessors act independently, and typically different sets of pairs are presented to different assessors, varying also their order. An assessor does not have the possibility to go back and check the answers she gave previously, let alone change any answer later.  Under such circumstances, often some answers given by an individual assessor, when considered afterwards jointly, do not satisfy logical transitivity of preferences \citep{tversky1969intransitivity}, that is, they may contain a pattern of the form $x \precc y\,,\,\, y \precc z $ but $z \precc x $. 
 On the other hand, neither are the answers given by an individual assessor independent, because conscientious assessors will generally try to follow some logic in their expressed preferences. 
 
Pair comparisons are preferred to ratings or full rankings of a set of items when there are many items to be compared, or when the relative differences between them are small: in both these cases assessors are unlikely to be able to inspect and compare all items jointly in order to perform a full ranking. A pairwise comparison test is then often preferred, and sometimes it is the only possible experimental procedure \citep{agresti}.

In this paper we consider pairwise comparison data coming from an experiment where each assessor was asked to hear a series of two different abstract  sounds,  and to tell which one was perceived as more human. 
Each subject only performed a limited number of comparisons, leading to sparse data, where not all pairs of sounds were compared by each assessor.
The results of this test are relevant for musicologists, composers and sound designers, whose aim is to understand how human performance expression can be communicated through spatial audio, leading to computer generated sounds appearing more life-like. Although every sound can be regarded as `spatial' in that sound waves propagate through space, the term `spatial audio' is here used to describe the way sound captures the physical movement in 3-D needed to produce it.
The cohort of listeners who took part in the experiment had varying backgrounds, ranging from musicologists to non-specialized university students. Therefore we expected listeners to cluster into groups, sharing different opinions about the degree of human causation behind sounds.
In addition to the grouping of the listeners around a shared consensus ranking of the ``humanness of sounds'', we were interested in studying the association between individual listeners' rankings and their own musical experience or musical background.
This application is described in detail in Section \ref{soni}.

Non-transitivity can arise for many reasons, for example assessors' inattentiveness, uncertainty in their preferences,  and actual confusion, even when one specific criterion for ranking is used.
These situations are so common that most pairwise comparison data are in fact non-transitive at the individual level, thus creating  a need for methods able to predict individual preferences from pairwise choices that lack logical transitivity, and only involve a very limited number of pair comparisons. Notice that the kind of non-transitivity that we consider in this paper regards only the individual level preferences. A different type of non-transitivity arises when aggregating preferences across assessors, as under Condorcet \citep{condorcet} or Borda \citep{de1781memoire} voting rules. 

We propose a new method for the analysis of pairwise comparison data that may contain non-transitive individual pairwise comparisons. The method is based on the classical Mallows rank model \citep{Mallows1957} and builds on its recent extension introduced by \citet{vitelli17}. Given pairwise data provided by a collection of individual assessors, the method outputs Monte Carlo samples from the joint posterior distribution for the individual full rankings of all items and an assumed shared consensus ranking between them. In Section \ref{mistura}, this hierarchical structure is further relaxed by introducing a mixture model allowing for clustering of the assessors.

The key ingredient, compared to \citet{vitelli17}, is to add to the model hierarchy one more layer of latent variables, accounting for the possibility that the assessors can make mistakes. By a mistake we mean that the order from an assigned pairwise comparison is reported in a way which is not consistent with the assessor's own `true' full ranking, whose existence is assumed in the model. The rationale behind our model can be explained as follows. In an ideal situation an assessor would be fully conscious of her preference ordering of all items, and then simply report the consequent ordering each time a pairwise comparison is requested. More realistically, however, she becomes aware of her potential ranking of the items only progressively in time as more pairs are presented to her for comparison. Then it becomes increasingly more difficult to remember exactly what items had been shown earlier and how they had been ordered, with the consequence that reporting results from pairwise comparisons that do not respect transitivity becomes more and more likely. Under such circumstances, particularly when the number of items is larger, the pair comparison data will almost inevitably contain some answers which do not satisfy the requirement of logical transitivity with the rest. Technical errors, such as mistakes in typing, or concentration errors may also occur. 

To describe such imperfections in the assessments, we introduce two alternative variants (described in Sections \ref{BM}
and \ref{LM}) of the probabilistic model for mistakes:
\begin{enumerate}
\item The probability of making a mistake is constant, independent of the pairs being assessed, and independent of all other comparisons made by the same assessor. 
\item The probability of making a mistake depends on the items being compared, and is higher for pairs which are more similar to each other. 
\end{enumerate}

The literature on inferential models for non-transitive pair data arising at the individual level is limited and discussed in Section \ref{related}. As far as we can see, the present paper stands out as the only approach to non-transitive pair data, when the individual hidden rankings are of interest, the same pairs are not repeatedly assessed by each assessor and are few, and a Bayesian approach is of interest.
One important feature of our Mallows model is the possibility to choose, for the considered specific application, an appropriate distance function. Some problems require a distance able to measure only the disorder in the given domain, while in others a distance more suited for learning preferences in a population would be preferred. In the former case, the Cayley distance \citep{cdist} would be a natural choice, while Kendall  \citep{kdist} and footrule \citep{sdist} would have advantages in the latter. For instance, consider the two rankings $\bm{\sigma}_1=(1,2,3,4,5)$ and $\bm{\sigma}_2=(5,2,3,4,1)$, where the top and bottom elements of $\bm{\sigma}_1$ are reversed in  $\bm{\sigma}_2$. The normalized Cayley distance between $\bm{\sigma}_1$ and  $\bm{\sigma}_2$ is 0.25, while the normalized Kendall distance is 0.7. If $\bm{\sigma}_1$ and $\bm{\sigma}_2$ represent the rankings of two assessors of five movies, Kendall's distance may be more appropriate, as these rankings represent very different profiles: one of the two assessors likes most the movie that the other assessor likes least, and vice versa. However, $\bm{\sigma}_1$ and $\bm{\sigma}_2$ differ by a unique translocation: if they represent genomes, we could consider these rankings as very similar and be more eager to use the Cayley distance as metric in the Mallows model. 
For a detailed description of the distances mentioned and of their properties we refer to \citet[][Chapter 6]{Diaconis1988}.

Our method provides the posterior distribution of the consensus ranking, as well as  the posterior distribution of the latent individual rankings for each assessor. The consensus ranking can be seen as a model-based Bayesian aggregation of individual preferences of a group of assessors. It is analogous to the quantities which are usually of interest in the \emph{rank aggregation} literature \citep{negahban2012iterative, dwork2001rank, kenyon2007rank, rajkumar2015ranking}. 
The estimated posterior distributions of the individual rankings can  be of great interest, for example, when performing personalized recommendations, or in studying how individual preferences change with assessor related characteristics.

This paper is organized as follows. In Section \ref{soni} we describe the application which motivated this study, and then in Section \ref{model} we present our model for the statistical analysis of the consequent data.
Numerical inference is based on a Markov Chain Monte Carlo algorithm, outlined in Section \ref{algo}.
Section 5 gives a short overview on other methods for the description and analysis of pairwise comparisons.
Section \ref{simus} is devoted to simulations, while in Section \ref{res_soni} we apply our method to the sound data, showing that the model identifies meaningful clusters of listeners, with similar perception of electroacoustic sounds. 
Finally, in Section \ref{concl} we summarize the  contributions of this paper.

\section{\emph{Acousmatic} music experiment}\label{soni}

\emph{Acousmatic} music is a type of electronic music composed for presentation using loudspeakers, as opposed to live or video recorded performance.
 The composer  manipulates digitally recorded sounds, so that the cause of the sound, being a musical instrument or any other sound making system, remains hidden. Indeed, when sounds are played over loudspeakers there are no visual cues to help listeners understand how the sounds were made. 
On the other hand, when we hear the sound of musical instruments or sounds from our everyday environment,  we are able to  recognize their cause, since in visual music we obtain the information that indicates the sounding object, i.e. its causation. Since the advent of recording technology, abstract sounds  (that is, sounds transformed with computer tools) have been used  in much of the sound-world we experience over the Internet, TV and film.

The question of interest is related to the ability of listeners to identify the presence of human causation through the spatial behavior of abstract sounds. \emph{Spatial} in this context describes the fact that the causation of sound happens as an action in 3-D space. The starting point for the experiment was a high-speed motion tracking recording of the physical movement used to produce one selected sound: a cellist bowing a down-bow chord. Features of this 3-D movement were successively subtracted, resulting in a series of 12 motion data-sets of varying proximity to the original. The motion data were then made audible by a process called parameter-mapping sonification \citep{grond2011}, where parameters in the data are mapped to parameters controlling computer generated sound.
The mapping rules are chosen to draw on our everyday perception of spatial motion, which involves not only absolute 3-D spatial location but in addition changes in volume, intensity and pitch, correlated with changes in proximity and speed. In other words, listeners heard the physical spatial motion through sonification,  rather than hearing the sound that the motion created, which, in this instance, was the sound of the cello.
Testing how listeners perceive a sound for which we lack a clear and commonly understood descriptive vocabulary is problematic. Therefore pair comparisons is the most appropriate design.  

\subsection{Pair comparison experiment}
The total number of stimuli was $12$. 
Test stimulus 1 (S1) was designed to most clearly sonify all features of the data. 
Each of the other 11 test stimuli were sonified by modifying one or more features of the data. This involved removing pitch and volume variation, flattening directional changes in the motion, or slowing the overall motion speed 
(as summarized in Table \ref{tab:soundsIntro}).
\begin{table}[h!]
\centering
\caption{Summary of the test sounds.}
\label{tab:soundsIntro}
\begin{tabular}{|p{0.35cm}p{11.5cm}|}
\hline
\textbf{S1}:  &  Pitch, volume, grain duration and spatial variations at their most dynamic ranges.                                                                                                                   \\ \hline
\textbf{S2}:  & Spatial motion occurring in front.                                                                                                                                                                                                                       \\ \hline
\textbf{S3}:  & Played in mono over one speaker direct-front.                                                                          \\ \hline
\textbf{S4:}  & Partial flattening of 3-D spatial variation leaving the main direction changes.   \\ \hline
\textbf{S5:}  & Total flattening of 3-D spatial variation leaving the main direction changes. \\ \hline
\textbf{S6:}  & Removal of volume variation.                                                                                                                                                                                                                                        \\ \hline
\textbf{S7:}  & Removal of pitch variation.                                                                                                                                                                                                                                          \\ \hline
\textbf{S8:}  & Removal of pitch and volume variation.                                                                                                                                                                                                                              \\ \hline
\textbf{S9:}  & Partial flattening of 3-D spatial variation; removal of pitch and volume variation.                                                                                                                                                                                                                              \\ \hline
\textbf{S10:} & Total flattening of 3-D spatial variation; removal of pitch and volume variation.                                                                                                                                                                                                                               \\ \hline
\textbf{S11:} & S1 played 30\% slower.                                                                                                                                                                                                                                                                     \\ \hline
\textbf{S12:} & S1 played 50\% slower.                                                                                                                                                                                                                                                        \\ \hline
\end{tabular}
\end{table}

Each of the 46 listeners involved in the experiment was exposed to $30$ pairs of these sounds, which is ca. 45\% of the total number of possible pairs of 12 stimuli.  The pairs were chosen randomly, without repetitions, and independently for each assessor. The items in each pair were played in randomized order.  

Listeners were then asked  to indicate, for each pair, which of the two stimuli most evoked a sensation of human physical movement of any kind, to follow their feelings, rather than imagining to watch a performance. The listeners were not told that the source motion stemmed from a cellist, nor were they asked to identify a specific human spatial movement. 
Each listener carried out the test  sitting centrally to the loudspeaker array. Prior to the experiment, listeners were presented with a short training session of three sounds not used in the test sequence. When the experiment began, the pairs of sounds were played sequentially, listeners noted their answers on a chart, selecting the first or the second from each pair of unlabeled sounds, and were requested to always make a choice even if they found it difficult to decide. 
If needed, they could ask to hear a test pair for a second time. 
At the end, they were asked to  complete two questionnaires, the aim of which was to assign a Musical Sophistication Index score  (MSI) and a rating of Spatial Audio awareness (SAA) to all the listeners.
The MSI used was the Ollen musical sophistication index \citep{ollen2006criterion}, which is an online survey that tests the validity of 29 indicators of musical sophistication. The SAA index consisted of five questions as indicators of how aware listeners were of spatial audio regardless of musical background. Such a test did not  exist in the literature, and was custom designed for the experiment.

The choice to rely on a pairwise comparison experiment is crucially   based on the listeners' lack of experience with abstract sounds. It is  easier for the participants to compare two sounds, rather than to be exposed to several, which could create confusion. 
The experiment, indeed, was difficult as expected: 37 listeners (80\%) reported non-transitivities in their pair comparisons, only 9 out of 46 listeners were able to stay consistent with themselves.

A complete description of the background, hypotheses, experimental setup, and discussion of results in \citet{natasha2017}.

\section{Bayesian Mallows models for non-transitive pairwise comparisons}\label{model}
We  consider the situation where  $N$  assessors independently express their preferences between pairs of the $n$ items in $\mathcal{O}=\{O_1,...,O_n\}$.
In many situations of practical interest the assessors do not decide on the set of pairs to be considered, which are instead assigned to them by an external authority. 
In this paper we decided not to model the way in which the pairs are chosen, and simply assume that each assessor $j$ receives a different subset $\mathcal{C}_j=\{\mathcal{C}_{j1},...,\mathcal{C}_{jM_j}\}$ of $M_j\leq n(n-1)/2$ random pairs.
Let  $\mathcal{B}_{j}=\{\mathcal{B}_{j1},...,\mathcal{B}_{jM_j}\}$  be the  set of  pairwise preferences given by assessor $j$, where $\mathcal{B}_{jm}$ is the order that assessor $j$ assigned to the pair $\mathcal{C}_{jm}$. 
For example, if  $\mathcal{C}_{jm}=\{O_{m_1},O_{m_2}\}$, it could be that  
$\mathcal{B}_{jm} =(O_{m_1}\precc O_{m_2})$, ${m_1}, {m_2}\in\{1,...,n\}$, meaning that  item $O_{m_1}$ is preferred to item $O_{m_2}$. Such data can be incomplete since not all items, nor pairs, are always handled by each assessor. We assume no ties in the data, that is, assessors are forced to express their preference for all pairs in the list $\mathcal{C}_{j}$ assigned to them, and indifference is not permitted.

We denote a generic ranking by $\bm{r}=(r_{1},...,r_{n})\in\mathcal{P}_n$, where $r_{i}\in\{1,...n\}$ is the rank of item $O_i$ (the most preferred item has rank $r_{i}=1$), and $\mathcal{P}_n$  is the space of $n$-dimensional permutations.
A widely used distance-based  family of distributions for ranks is the Mallows model \citep{Mallows1957,Diaconis1988}.
According to the Mallows model, the probability density of a given ranking  $\bm{r}=(r_1,...,r_n)$, here denoted by $\mathcal{M}al(\bm{\rho}, \alpha),$ is given by
 \begin{equation}\label{Mallows}
 f_{\bm{R}}(\bm{r}\,|\,\alpha,\bm{\rho}) :=\frac{\exp[-\frac{\alpha}{n}d(\bm{r}, \bm{\rho})]}{Z_n(\alpha)}\mathbb{1}_ {\mathcal{P}_n}(\bm{r}).
 \end{equation}
In \eqref{Mallows}, $\bm{\rho}\!\in\!\mathcal{P}_n$ is  the location parameter representing the shared consensus ranking, $\alpha\!>\!0$  is the scale parameter measuring the concentration of the data around $\bm{\rho}$, and $d(\cdot, \cdot) $ is a distance function between two $n-$dimensional permutations that satisfies right-invariance \citep{Diaconis1988}, i.e., $d(\bm{r}, \bm{\rho})=d(\bm{r}\circ\bm{r}', \bm{\rho}\circ\bm{r}')$, $\forall \bm{r},\bm{r}',\bm{\rho}\in\mathcal{P}_n$, where  $\bm{\rho}\circ\bm{r}'=\bm{\rho}_{\bm{r}'}=(\rho_{r'_1},...,\rho_{r'_n})$.
Right-invariance is crucial  since from this property it follows that the partition function of  \eqref{Mallows} does not depend on the location parameter, and can then be written as \mbox{$Z_n(\alpha)=\sum_{\bm{r}\in\mathcal{P}_n} \exp\left\{-\frac{\alpha}{n}d(\bm{r}, \bm{1}_n)\right\}$}, where  $\bm{1}_n=(1,...,n)$ (see for example \citet{mukherjee2016}).
When the distance function in \eqref{Mallows} is chosen to be the Kendall, the Cayley, or the Hamming distance, the partition function of the Mallows model is available in closed form \citep{Fligner1986}. For this reason, most of the work on the Mallows was limited to these distances (see, for example, \citet{Fligner1986}, \citet{Lu2015}, \citet{irurozki2016sampling, irurozki2014Ham}).
The Mallows with other distance functions was less treated because of  its computational complexity.
Recently, \citet{vitelli17} gave a procedure to compute $Z_n(\alpha)$ when the footrule and Spearman distances are used, either exactly (up to some moderate values of $n$), or approximated through an Importance Sampling technique. 
The authors set the original Mallows model  in a Bayesian framework, also allowing for data in the form of transitive pairwise comparisons. We generalize their model (described in Section 4.2 of \citet{vitelli17}) to handle non-transitive pairwise comparisons.

The main assumption is that each assessor $j$ has a personal latent  ranking, $\bm{R}_j=(R_{j1},...,R_{jn})\in\mathcal{P}_n$, distributed according to the Mallows density  \eqref{Mallows}, 
$\bm{R}_1,...,\bm{R}_N | \bm{\rho},\alpha \iidsim \mathcal{M}al(\bm{\rho},\alpha).$
We model the situation where each assessor $j$, when announcing her preferences, matches the items under comparison with her latent ranking $\bm{R}_j$.
Then, if the assessor is consistent with   $\bm{R}_j$, the pairwise orderings in $\mathcal{B}_{j}$ are induced by $\bm{R}_j$ according to: 
\begin{equation}\label{pref}\begin{split}(O_{m_1}\precc O_{m_2}) \text{     }\iff\text{     } R_{j{m_1}}<{R}_{j{m_2}} ,
\end{split}\end{equation}
where $R_{jm_i}$ denotes the rank of item $O_{m_i}$ in $\bm{R}_j$.
In this case  the set of pairwise orderings $\mathcal{B}_{j}$ contains  only mutually compatible (a.k.a. transitive) preferences, since the preferences are induced from a complete ranking in $\mathcal{P}_n$ that, by definition, is transitive. 
The transitive closure of a set of pairwise preferences, denoted by $\text{tc}(\mathcal{B}_{j})$, is the smallest set that consistently extends the original preference set: it is defined as the set union of $\mathcal{B}_j$  and all pairwise preferences  that are not explicitly given but are induced from $\mathcal{B}_j$ by transitivity.
In this case it is possible to first compute $\text{tc}(\mathcal{B}_{j})$, and second, to make inference on the posterior distribution of the Mallows parameters by integrating out all the rankings $\bm{{r}}\in\mathcal{P}_n$ that are compatible with the transitive closure of the preference sets, denoted  by $\bm{{r}}\leftarrow\text{tc}(\mathcal{B}_{j})$,
\begin{equation}\label{cons}
\pi(\alpha,\bm{\rho}| \,\mathcal{B}_{1},...,\mathcal{B}_{N})\propto
\pi(\alpha)\pi(\bm{\rho})\prod_{j=1}^N\left[\sum_{\bm{{r}}\leftarrow\text{tc}(\mathcal{B}_{j})} f_{\bm{R}_j}( \bm{{r}}|\,\alpha,\bm{\rho})
\right].\end{equation}
This setting was described in \citet{vitelli17}, Section 4.2.

If  the assessor is not fully consistent with her latent ranking, the pairwise orderings in $\mathcal{B}_{j}$ may not be mutually compatible. In such a case  the transitive closure may not exist and the previous procedure cannot be followed. Therefore a model able to account for non-transitive patterns in the data is needed in this setting.

We propose a probabilistic strategy based on the assumption that non-transitivities are due to mistakes in deriving the pair order from the latent raking  $\bm{R}_j$. 
The likelihood assumed for a set of preferences $\mathcal{B}_{j}$ (analogous to the summation of eq. (\ref{cons})) is
\begin{equation}\label{sum1}
f(\mathcal{B}_{j}|\alpha,\bm{\rho})=\sum_{\bm{{r}}\in\mathcal{P}_n}f(\mathcal{B}_{j}, \bm{R}_j=\bm{r}|\alpha,\bm{\rho})=\sum_{\bm{{r}}\in\mathcal{P}_n}f_{\bm{R}_j}(\bm{{r}}|\,\alpha,\bm{\rho})f(\mathcal{B}_{j}|\bm{R}_j=\bm{{r}}) ,
\end{equation}
\noindent where $f(\mathcal{B}_{j}|\bm{R}_j=\bm{{r}})$  is  the probability of ordering the pairs in $\mathcal{C}_j$ as in  ${\mathcal{B}}_{j}$ (possibly generating non-transitivities), when the latent ranking for assessor $j$ is $\bm{R}_j=\bm{{r}}$. It can therefore be seen as forming the error model in this context, which will be specified below.
The joint posterior of the model parameters is then:
\begin{equation}\label{incons}\begin{split}
\pi(\alpha,\bm{\rho}| \,\mathcal{B}_{1},...,\mathcal{B}_{N})\propto\,\pi(\alpha)&\pi(\bm{\rho}) \prod_{j=1}^N\left[\sum_{{\bm{r}}\in\mathcal{P}_n}f_{\bm{R}_j}(\bm{{r}}|\,\alpha,\bm{\rho})f(\mathcal{B}_{j}|\bm{R}_j=\bm{{r}})
\right].\end{split}\end{equation}
In this paper we have assumed a gamma prior, $\pi(\alpha)=\frac{\lambda^\gamma}{\Gamma(\gamma)}\alpha^{\gamma-1} e^{-\lambda\alpha}\mathbb{1}_{\mathbb{R}^+}(\alpha)$, for $\alpha$, and the uniform prior on $\mathcal{P}_n$, $\pi(\bm{\rho})=\frac{\mathbb{1}_{\mathcal{P}_n}(\bm{\rho})}{n!}$, for $\bm{\rho}$.

This strategy is able to recover possible linear orderings close (in terms of some given distance) to the non-transitive sets of  preferences. 
We developed two basic models for the probability of making a mistake: the Bernoulli  model (BM) and the Logistic  model (LM). BM assumes that  non-transitivities arise from random mistakes while LM assumes that non-transitivities arise from mistakes due to difficulty in ordering similar items.

\subsection{Bernoulli model (BM)} \label{BM}
Assume that the pairwise comparisons given by an assessor are conditionally independent given her latent ranking $\bm{R}_{j}$,
\begin{equation}\begin{split}f(\mathcal{B}_{j}|\bm{R}_j=\bm{r})=\prod_{m=1}^{M_j}f(\mathcal{B}_{jm}|\bm{R}_j=\bm{r})
 .\end{split}\label{bm1}\end{equation}
We define here a function of a given comparison $\mathcal{B}_{jm} =(O_{m_1}\precc O_{m_2})$, and of a given ranking $\bm{r}=(r_1,...,r_n)\in\mathcal{P}_n$, 
$\text{g}(\mathcal{B}_{jm},\bm{r})= \mathbb{1}(r_{m_1}>r_{m_2})$, where $m_1$ is the index of the preferred item $O_{m_1}$ in the $m$-th comparison $\mathcal{B}_{jm}$ of assessor $j$, and $m_2$ is the index of the less preferred item. Thus $\text{g}(\mathcal{B}_{jm},\bm{r})=1$ if the preference  order of  $\mathcal{B}_{jm}$ contradicts with that implied by the ranking $\bm{r}$  (in the sense of eq. (\ref{pref})).

We then assume the following Bernoulli type model for modeling the probability that an assessor $j$ makes a mistake in a given pairwise comparison $\mathcal{B}_{jm}$, that is the probability that she reverses the true latent preference implied by her latent ranking $\bm{R}_j$:
\begin{equation*}\begin{split}\mathbb{P}(\mathcal{B}_{jm}\; \text{mistake} \,|\, \theta, \bm{R}_{j}=\bm{r})=\mathbb{P}(\text{g}(\mathcal{B}_{jm},\bm{r})=1\,|\, \theta, \bm{R}_{j}=\bm{r})=\theta,\quad\theta\in[0,0.5) .\end{split}\end{equation*}
Eq. (\ref{bm1}) is then  given by 
{\begin{equation*}\begin{split}
f(\mathcal{B}_{j}\,| \,\theta,\bm{R}_j=\bm{r})= \left(\frac{\theta}{1-\theta}\right)^{\sum_{m=1}^{M_j}\text{g}(\mathcal{B}_{jm},\bm{r})}(1-\theta)^{M_j}\,.
\end{split}\end{equation*}}
We assign to $\theta$ the truncated Beta distribution on the interval $[0,0.5)$ as prior, with given hyperparameters $\kappa_1$ and $\kappa_2$: $\pi(\theta)\propto\theta^{\kappa_1-1}(1-\theta)^{\kappa_2 -1}\mathbb{1}_{[0,0.5)}(\theta)$, conjugate to the Bernoulli model. We choose the truncated Beta mainly for identification purposes, but this choice is also motivated by the fact that we want  to force the probability of making a mistake to be less than 0.5.

Let  $\mathcal{B}_{1:N}$ be a shorthand for $\mathcal{B}_{1},...,\mathcal{B}_{N}$, and $\bm{R}_{1:N}$  for $\bm{R}_{1},...,\bm{R}_{N}$. \\The posterior density of the model parameters, defined on the support \\
$S=
\mathbb{1}\left(\{\alpha>0\}\cap\{\bm{\rho}\in\mathcal{P}_n\}\cap\{\bm{R}_{j}\in\mathcal{P}_n\}_{j=1}^N\cap\{0\leq\theta<0.5\}\right)$,
 has the following form,
\begin{equation}\begin{split}\pi(\alpha,\bm{\rho}, \theta|\mathcal{B}_{1:N})&\!\propto\! \pi(\alpha)\pi(\bm{\rho})\pi(\theta)\prod_{j=1}^N \left[\sum_{\bm{r}\in\mathcal{P}_n}f_{\bm{R}_j}({\bm{r}} |\alpha,\bm{\rho})f({\mathcal{B}}_{j}|\theta,\bm{R}_j={\bm{r}})\right].\end{split}\label{postber}
\end{equation}

We sample from the density of eq. \eqref{postber} through an augmented sampling scheme, by first updating $\alpha,\bm{\rho}$ and $\theta$ given $\mathcal{B}_{1:N}$ and $\bm{R}_{1:N}$, and then updating $\bm{R}_{1:N}$ given  $\alpha,\bm{\rho},\theta$ and $\mathcal{B}_{1:N}$. The former step is {\color{black}performed} by using the conditional density
\begin{equation}\begin{split}
\pi(\alpha,&\bm{\rho}, \theta|\mathcal{B}_{1:N}, \bm{R}_{1:N}) =\alpha^{\gamma-1}e^{-\alpha\left(\lambda+\frac{1}{n}\sum_{j=1}^Nd(\bm{R}_j,\bm{\rho})\right)
-N\ln[Z_n(\alpha)]}\\
\cdot\,&\left(\frac{\theta}{1-\theta}\right)^{\kappa_1-1+\sum_{j=1}^N\sum_{m=1}^{M_j}\text{g}(\mathcal{B}_{jm},\bm{R}_j)}
(1-\theta)^{\kappa_2+\kappa_1-2+\sum_{j=1}^NM_j}.
\end{split}
\label{postber1}
\end{equation}
The second step is performed by using the density
\begin{equation}\begin{split}
\pi(\bm{R}_{1:N}|&\alpha,\bm{\rho},\theta,\mathcal{B}_{1:N})\propto\pi(\bm{R}_{1:N}|\alpha,\bm{\rho})\pi(\mathcal{B}_{1:N}|\theta,\bm{R}_{1:N}) =\\
=&\frac{e^{-\frac{\alpha}{n}\sum_{j=1}^Nd(\bm{R}_j,\bm{\rho})}}{[Z_n(\alpha)]^N}\left(\frac{\theta}{1-\theta}\right)^{\sum_{j=1}^N\sum_{m=1}^{M_j}\text{g}(\mathcal{B}_{jm},\bm{R}_j)}(1-\theta)^{\sum_{j=1}^NM_j}.
\end{split}
\label{postber2}
\end{equation}

\subsection{Logistic model (LM)}\label{LM}
The idea {\color{black}behind} the logistic model for mistakes is that an assessor $j$ is more likely to be confused (and consequently to make a mistake) if two items in a pair are  more similar according to her latent rank vector $\bm{R}_j$. 
We assume the following logistic type model for the probability of making a mistake in a given pairwise comparison 
{\begin{equation*}\begin{split}\text{logit}\,\mathbb{P}\left(\mathcal{B}_{jm}\;\text{mistake}\Big|{ \bm{R}}_{j},\beta_0,\beta_1\right)=-\beta_0-\beta_1\frac{d_{\bm{R}_j,m}-1}{n-2} ,
\end{split}\end{equation*}}
where  $d_{\bm{R}_j,m}$ is the  $\ell_1$ distance of the ranks of the two items under comparison in $\mathcal{B}_{jm}$,  according to $\bm{R}_j$: if $\mathcal{B}_{jm}=(O_{m_1}\precc O_{m_2})$, then 
$d_{\bm{R}_j,m}=|R_{jm_1}- R_{jm_2}|$.
We assume that $\beta_1$ and  $\beta_0$ are a priori independent and  distributed according to a gamma prior, $\beta_1\sim\Gamma(\lambda_{11},\lambda_{12})$, and $\beta_0\sim\Gamma(\lambda_{01},\lambda_{02})$.
These choices are motivated by the fact that we want to model a negative dependence between the distance of the items and the probability of making a mistake ($\beta_1>0$), and second, we want  to force the probability of making a mistake when the items have ranks differing by  1 to be less than 0.5 ($\beta_0>0$).   
The posterior density of the model, defined on {\color{black}the support} $S=
\mathbb{1}\left(\{\alpha>0\}\cap\{\bm{\rho}\in\mathcal{P}_n\}\cap\{\bm{R}_{1:N}\in\mathcal{P}_n\}\cap\{\beta_1>0\}\cap\left\{\beta_0>0\right\}\right)$,
 is then
 \begin{equation}\begin{split}\pi(\alpha,\bm{\rho}, \beta_0, \beta_1|\mathcal{B}_{1:N})&\propto \pi(\beta_0)\pi(\beta_1)\pi(\bm{\rho})\pi(\alpha)\\\cdot&\prod_{j=1}^N \left[\sum_{\bm{r}\in\mathcal{P}_n}f_{\bm{R}_j}({\bm{r}} |\alpha,\bm{\rho})f({\mathcal{B}}_{j}|\beta_0, \beta_1,\bm{R}_j={\bm{r}})\right].\end{split}\label{postlogi}
\end{equation} 
Analogously to eq. \eqref{postber}, we sample from the posterior of eq. \eqref{postlogi} by first updating $\alpha,\bm{\rho}, \beta_0$ and $\beta_1$, given $\mathcal{B}_{1:N}$ and $\bm{R}_{1:N}$, i.e. from 

\begin{equation}\label{postlm1}\begin{split}
\pi&(\alpha,\bm{\rho}, \beta_0, \beta_1|\mathcal{B}_{1:N}, \bm{R}_{1:N})\propto\frac{\alpha^{\gamma-1}\beta_0^{\lambda_{01}-1}\beta_1^{\lambda_{11}-1}}{\prod_{j=1}^N\prod_{m=1}^{M_j}\left[1+e^{-\beta_0-\beta_1\frac{d_{\bm{{R}}_j, m}-1}{n-2}}\right]}\\
&\cdot e^{-\alpha\left(\lambda +\frac{1}{n}\sum_{j=1}^Nd(\bm{R}_j,\bm{\rho})\right)-N\ln[Z_n(\alpha)] -\beta_0\left[\lambda_{02}+\sum_{j=1}^N\sum_{m=1}^{M_j}\text{g}(\mathcal{B}_{jm},\bm{R}_j)\right]}\\  
&\cdot e^{-\beta_1\left[\lambda_{12}+\frac{1}{n-2}\sum_{j=1}^N\sum_{m=1}^{M_j}{\text{g}(\mathcal{B}_{jm},\bm{R}_j)}(d_{\bm{{R}}_j, m}-1)\right]}.
\end{split}\end{equation}

Secondly, we update 
 $\bm{R}_{1:N}$, given $\alpha,\bm{\rho}, \beta_0, \beta_1$ and $\mathcal{B}_{1:N}$, from 
\begin{equation}\label{postlm2}\begin{split}
\pi&(\bm{R}_{1:N}|\alpha,\bm{\rho}, \beta_0, \beta_1,\mathcal{B}_{1:N})\propto
\pi(\bm{R}_{1:N}|\alpha,\bm{\rho})\pi(\mathcal{B}_{1:N}|\beta_0,\beta_1,\bm{R}_{1:N}) \propto\\
&\propto e^{-\frac{\alpha}{n}\sum_{j=1}^Nd(\bm{R}_j,\bm{\rho})-N\ln[Z_n(\alpha)]-
\beta_0\sum_{j=1}^N\sum_{m=1}^{M_j}\text{g}(\mathcal{B}_{jm},\bm{R}_j)}\\
&\cdot e^{-\frac{\beta_1}{n-2}\sum_{j=1}^N\sum_{m=1}^{M_j}{\text{g}(\mathcal{B}_{jm},\bm{R}_j)}(d_{\bm{{R}}_j, m}-1)}\left[\prod_{j=1}^N\prod_{m=1}^{M_j}\left(1+e^{-\beta_0-\beta_1\frac{d_{\bm{{R}}_j, m}-1}{n-2}}\right)\right]^{-1}.
\end{split}\end{equation}

\subsection{Clustering non-transitive assessors}\label{mistura}
So far we assumed that a unique consensus ranking was shared by all assessors. Since in many situations this assumption is unrealistic, we  allow for clustering the assessors into separate subsets, each sharing a consensus ranking of the items.
We propose a mixture model generalization of the Bernoulli model of Section \ref{BM} to deal with  heterogeneous assessors expressing pairwise preferences with mistakes.  

Let $z_1,...,z_N\in \{1,...,G\}$ be the class labels indicating how individual assessors are assigned to one of the $G$ clusters. Each cluster is described by a different pair of Mallows parameters $(\alpha_g, \bm{\rho}_g)$, $g=1,...,G$, {\color{black}so that} the likelihood has  the following form:
\begin{equation*}\begin{split}\label{lik_mixture}
f(\mathcal{B}_{1:N}|\alpha_{1:G},\bm{\rho}_{1:G},\theta,\eta_{1:G}, z_{1:N})\!=\!\prod_{j=1}^N\left\{\sum_{{\bm{r}}\in\mathcal{P}_n}\!f_{\bm{R}_j}({\bm{r}}|\alpha_{z_j},\bm{\rho}_{z_j})f(\mathcal{B}_{j}|\theta,{\bm{R}_j}={\bm{r}}\!)\right\},
\end{split}
\end{equation*}
where 
{$$f_{\bm{R}_j}({\bm{r}}|\alpha_{z_j},\bm{\rho}_{z_j})= \frac{\mathbb{1}_{P_n}({\bm{r}})}{Z_n(\alpha_{z_j})} \exp\left\{-\frac{\alpha_{z_j}}{n}d(\bm{r}_,\bm{\rho}_{z_j})\right\}\,.$$}
We assume that the cluster labels are a priori conditionally independent given the mixing parameters of the clusters, $\eta_1,...,\eta_G$,  and distributed according to a categorical distribution
$$\pi(z_1,...,z_N|\eta_1,...,\eta_G)\propto\prod_{j=1}^N\eta_{z_j}=\prod_{j=1}^N\prod_{g =1}^G\eta_{g}^{\mathbb{1}_g(z_j)}\,\,,$$
where $\eta_g\,\geq\,\, 0$, $\forall g =1,...,G$ and $\sum_g\eta_g=1$. Finally we assign to $\eta_1,...,\eta_G$ the Dirichlet density with parameter $\chi$.
These choices lead to the following posterior density,
\begin{equation}\begin{split}\label{postMixt}
\pi(\alpha_{1:G},\bm{\rho}_{1:G},& \eta_{1:G}, \theta, z_{1:N}|\mathcal{B}_{1:N})\propto \pi(\theta)\prod_{g =1}^G\left[\pi(\alpha_{g})\pi(\bm{\rho}_{g})
\pi(\eta_{g})\right]\\
\cdot&\prod_{j=1}^N\left[\pi(z_{j}|\eta_{1:G})\sum_{{\bm{r}}\in\mathcal{P}_n}f_{\bm{R}_j}({\bm{r}}|\alpha_{z_j},\bm{\rho}_{z_j})f(\mathcal{B}_{j}|\theta,\bm{R}_j=\bm{r})\right]\,.
\end{split}
\end{equation}

Similarly to the homogeneous case, we then sample from the posterior of eq. \eqref{postMixt} by first updating $\alpha_{1:G}, \bm{\rho}_{1:G}, \eta_{1:G}, z_{1:N}$ and $\theta$, given $\mathcal{B}_{1:N}$ and $\bm{R}_{1:N}$, and then updating $\bm{R}_{1:N}$, given  $\alpha_{1:G},\bm{\rho}_{1:G}, \eta_{1:G}, z_{1:N}, \theta$ and $\mathcal{B}_{1:N}$. The former step is done by using the conditional density
\begin{equation}\begin{split}\label{postMixt1}
\pi(\alpha_{1:G},&\bm{\rho}_{1:G}, \eta_{1:G}, z_{1:N}, \theta|\mathcal{B}_{1:N}, \bm{R}_{1:N})\propto
\prod_{g =1}^G\left[\alpha_g^{\gamma-1}e^{-\lambda\alpha_g}\eta_g^{\xi-1+\sum_{j=1}^N\mathbb{1}_g(z_j)}\right]\,\\
&\cdot\left(\frac{\theta}{1-\theta}\right)^{\kappa_1-1+\sum_{j=1}^N\sum_{m=1}^{M_j}\text{g}(\mathcal{B}_{jm},\bm{R}_j)}
(1-\theta)^{\kappa_2+\kappa_1-2+\sum_{j=1}^NM_j}\,\\
&\cdot\prod_{j=1}^N\left[ \frac{e^{-\frac{\alpha_{z_j}}{n}d(\bm{R}_j,\bm{\rho}_{z_j})}}{Z_n(\alpha_{z_j})}
\,\right].
\end{split}\raisetag{-4\baselineskip}
\end{equation}

The second step is performed by using the density 
\begin{equation}\begin{split}\label{postMixt2}
\pi(\bm{R}_{1:N}|&\alpha_{1:G},\bm{\rho}_{1:G}, \eta_{1:G}, \theta, z_{1:N},\mathcal{B}_{1:N})\propto\\
\propto&\prod_{j=1}^N\left[\frac{e^{-\frac{\alpha_{z_j}}{n}d(\bm{R}_j,\bm{\rho}_{z_j})}}{Z_n(\alpha_{z_j})}\left(\frac{\theta}{1-\theta}\right)^{\sum_{m=1}^{M_j}\text{g}(\mathcal{B}_{jm},\bm{R}_j)}(1-\theta)^{M_j}\right].
\end{split}
\end{equation}

Since label switching is not  handled inside our MCMC, MCMC iterations are re-ordered after convergence has been achieved, by applying the algorithm of \citet{stephens2000dealing}. 

\vspace{0.5cm}
\section{MCMC for non-transitive pairwise preferences}\label{algo}
We develop a  Markov Chain Monte Carlo (MCMC) algorithm which, at convergence, samples from  the posterior density of eq. (\ref{postber}).
As explained in Section \ref{BM}, the MCMC iterates between two main steps:
\begin{enumerate}
\item\label{step1} Update $\alpha,\bm{\rho}$ and $\theta$ given $\mathcal{B}_{1:N}$ and $\bm{R}_{1:N}$ (using eq. \eqref{postber1}):
\begin{enumerate}
\item Metropolis update of $\bm{\rho}$
\item Metropolis update of  ${\alpha}$ 
\item Gibbs update of $\theta$
\end{enumerate}
\item Update $\bm{R}_{1:N}$ given $\alpha,\bm{\rho},\theta$ and $\mathcal{B}_{1:N}$ (using eq. \eqref{postber2}).
\end{enumerate}

In  step 1(a), we propose a new consensus ranking $\bm{\rho}^p$ according to a symmetric proposal which is centered around the current consensus ranking  $\bm{\rho}^t$.

\begin{mydef}{\texttt{Swap} proposal.}\label{swap}
At step $t$, denote the current version of the consensus ordering vector by ${\bm{x}}^{t}=(\bm{\rho}^{t})^{-1}$, which is the vector whose $n$ components are the items in $\mathcal{O}$ ordered from best to worst according to $\bm{\rho}^{t}$, i.e., $x^t_i=O_k \iff \rho^t_k=i$. Let $L^*\in\{1,..,n\}$. Sample uniformly an integer $l$ from $\{1,2,...,L^*\}$ and draw a random number $u$ uniformly in $\{1,2,...,n-l\}$. The proposal ${\bm{x}}^p$ has components
\begin{equation}{{x}}_{i}^p=
\begin{cases}
{{x}}_{i}^t& \mbox{if  } \,\,\,i\not=\{u,u+l\}\\
{{x}}_{u+l}^t& \mbox{if  } \,\,\,i=u\\
{{x}}_{u}^t&\mbox{if  } \,\,\,i=u+l\\
\end{cases}
\end{equation}
and  the proposed ranking  is ${\bm{\rho}}^p=({\bm{x}}^p)^{-1}$.
\end{mydef}

The parameter $L^*$ is the maximum allowed distance between the ranks of the swapped items, and is used for tuning the acceptance probability in the Metropolis-Hastings step.
The transition probability of the \texttt{Swap} proposal is symmetric, and given by $q({\bm{\rho}}^{{\color{black}p}}\rightarrow{\bm{\rho}}^t)=\frac{1}{L^*}\sum_{l=1}^{L^*}\frac{1}{n-l}\mathbb{1}(|{{\bm{\rho}}^{{\color{black}p}}-{\bm{\rho}}^t}|=2l)$.
The  ranking is then accepted with probability $\epsilon = \min\{1, a_{\bm{\rho}}\}$, where
$$\ln(a_{\bm{\rho}})= -\frac{\alpha}{n} \sum_{j=1}^N \left[d(\bm{R}_j,\bm{\rho}^p)-d(\bm{R}_j,\bm{\rho}^t)\right] .$$

In step 1(b) we propose $\alpha^p$ from a log-normal density $\ln\mathcal{N} (\ln(\alpha^t), \sigma_\alpha^2 ),$ and accept it with probability $\epsilon = \min\{1, a_{\alpha}\}$, where
$$\ln(a_{\alpha})=\gamma\left[\ln(\alpha^p/\alpha^t)\right]-\bigg[\lambda+\frac{1}{n}\sum_{j=1}^Nd(\bm{R}_j,\bm{\rho})\bigg](\alpha^p-\alpha^t)
-N\bigg[\ln[Z_n(\alpha^p)/Z_n(\alpha^t)]\bigg].$$
This acceptance probability takes into account the asymmetric transition probability of the chain, that results from the log-normal proposal. 
The  partition function $Z_n(\alpha)$ can be computed exactly or approximated by the importance sampling scheme proposed by \citet{vitelli17}, depending on the distance function chosen and on the number $n$ of items considered.

In step 1(c) we sample $\theta$ from the beta distribution, truncated  to the interval $[0,0.5)$, with updated hyper-parameters, 
$$\kappa'_{1}=\kappa_{1}+\sum_{j=1}^N\sum_{m=1}^{M_j}\text{g}(\mathcal{B}_{jm},{\bm{R}_j}),\quad\quad\kappa'_{2}=\kappa_{2}+\sum_{j=1}^N\sum_{m=1}^{M_j}[1-\text{g}(\mathcal{B}_{jm},{\bm{R}_j})].$$

Step 2 is a Metropolis-Hastings for the individual rankings. Here we exploit the fact that, {\color{black}when} fixing all other parameters and the data $\mathcal{B}_1,...,\mathcal{B}_N$, $\bm{R}_1,...,\bm{R}_N$  are conditionally independent, and that each $\bm{R}_j$ only depends on the corresponding data $\mathcal{B}_j$. 
We thus sample a proposed individual ranking $\bm{r}_j^p$ from the \texttt{Swap} proposal, separately for each $j=1,...,N$.
The \texttt{Swap} proposal is here advantageous because it perturbs locally not only the current individual ranking $\bm{r}_j^t$, but also the function $\text{g}(\mathcal{B}_{jm},\bm{r}_j^t)$.

 \vspace{0.2cm}
\textsc{Remark.}{ The  \texttt{Swap} {\color{black}proposal} always gives  a proposed individual ranking $\bm{r}_j^p\not=\bm{r}_j^{t}$. However, it may happen that  $\text{g}(\mathcal{B}_{jm},\bm{r}_j^p)=\text{g}(\mathcal{B}_{jm},\bm{r}_j^t)$, $\forall m=1,...,M_j$.}
 \vspace{0.2cm}

This  is important for what concerns the acceptance probability of ${\bm{r}}^p_{\color{black}j}$. 
If $\text{g}(\mathcal{B}_{jm},\bm{r}_j^p)=\text{g}(\mathcal{B}_{jm},\bm{r}_j^t)$, $\forall m=1,...,M_j$, the acceptance probability  depends only on the ratio of the Mallows likelihoods of $\bm{r}_j^p$  and $\bm{r}_j^{t}$, and is equal to $\epsilon=\min\{1,{a}_1\}$, where
$$\ln({a}_1)=-\frac{\alpha}{n}\left[d(\bm{r}_j^p,\bm{\rho})-d(\bm{r}_j^{t},\bm{\rho})\right].$$

If $\text{g}(\mathcal{B}_{jm},\bm{r}_j^p)\not=\text{g}(\mathcal{B}_{jm},\bm{r}_j^t)$ for some $m=1,...,M_j$, the acceptance probability depends also on the mistake model, and is equal to $\epsilon=\min\{1,{a}_2\}$ where
$$\ln({a}_2)=\ln({a}_1) + \sum_{m=1}^{M_j}\left[\text{g}(\mathcal{B}_{jm},\bm{r}_j^p)-\text{g}(\mathcal{B}_{jm},\bm{r}_j^t)\right]\ln\left[\theta/({1-\theta})\right].$$
\textsc{Example.} {\textit{To illustrate this step of the algorithm, suppose that an assessor expresses the following set of preferences, {\small$$\mathcal{B}_j=\{\! (O_2\precc O_1), (O_5\precc O_4), (O_5\precc O_3), (O_5\precc O_2), (O_5\precc O_1), (O_3\precc O_2), (O_1\precc O_3)\! \}.$$}
\noindent This set contains the non-transitive pattern  \mbox{$O_2\precc O_1\precc O_3\precc O_2$.}
 For the illustration, suppose that the current value  of the individual ranking vector is $\bm{r}_j^{t}=(5,4,3,2,1)$, which corresponds to the ordering vector $\bm{x}_j^{t}=(O_5, O_4, O_3, O_2, O_1)$, and for which $\sum_{m=1}^7\text{g}(\mathcal{B}_{jm},\bm{r}_j^p)=1$.
If we sample the proposal  $\bm{x}_j^p=(O_5, O_3, O_4, O_2, O_1)$, this gives $\text{g}(\mathcal{B}_{jm},\bm{r}_j^p)=\text{g}(\mathcal{B}_{jm},\bm{r}_j^t)$, $\forall m=1,...,7$, and $\bm{r}_j^p=(5,4,2,3,1)\not=\bm{r}_j^{t}$. However, 
if we  sample $\bm{x}_j^p=(O_4, O_5, O_3, O_2, O_1)$, then $\bm{r}_j^p=(5,4,3,1,2)\not=\bm{r}^{t}$ and also $\sum_{m=1}^7\text{g}(\mathcal{B}_{jm},\bm{r}_j^p)=2\not=\sum_{m=1}^7\text{g}(\mathcal{B}_{jm},\bm{r}_j^t)$ since, according to the sampled $\bm{r}_j^p$, the preference $O_5\precc O_4$ is reversed. \\}}

Appropriate convergence of the MCMC must in practice be checked by inspecting the trace plots of the parameters, and by monitoring for example the integrated autocorrelation.
In \ref{suppA} we explain in detail how the algorithm is adapted to the case of the logistic mistake model, and to the mixture extension.


\section{Other approaches to pairwise preference data}\label{related} 
In the classical Bradley-Terry model (BT) for pair comparisons  \citep{bradley1952rank} the probability that item $O_i$ is preferred to item $O_k$ is expressed as the ratio
\begin{equation}\label{BT}
\text{Pr}(O_i\prec O_k|\bm{\mu})=\frac{{\mu_i}}{{\mu_i}+{\mu_k}},
\end{equation}
where $\bm{\mu}=(\mu_1,\dots,\mu_n)$, is a vector of item-specific consensus ratings shared by all assessors, forming a linear scale of score parameters.  From this follows that the odds for $(O_i \prec O_k )$ against $(O_k \prec O_i )$ are given by $\mu_i/\mu_k$. In addition, it is assumed that all pairwise comparisons are conditionally independent given $\bm{\mu}$. Therefore, the likelihood expression of the BT model corresponding to data consisting of several pair comparisons is the product, across all considered pairs, of terms of the form \eqref{BT}. For this reason, all pairwise data, even when they may have come from a number of individual assessors, are effectively merged when performing inference on $\bm{\mu}$.  

The work of Bradley and Terry was preceded by two important earlier papers, by \citet{thurstone1927law} and \citet{zermelo29}. Thurstone considered a similar preference data context as BT, but the work was based on a Gaussian error model. \citet{zermelo29}, in contrast, proposed exactly the same model as Bradley and Terry, but it was presented as a statistical model for the results from a chess tournament, without the presence of individual assessors. After these pioneering works, several extensions of the basic BT model have been presented, mostly in the econometric and psychometric literature. Often these papers apply the logarithmic transformation $u_i = \log \mu_i$ of the parameters, with the effect that the probabilities \eqref{BT} get the familiar logistic form. The logit of the odds for $(O_i \prec O_k )$ against $(O_k \prec O_i )$ is then equal to the contrast $u_i - u_k$ between the corresponding logarithmic scores. Extensions to regression models that account for the influence of item specific covariates on the comparison results are then readily available; for more comments on this, see below.  

Data generated from the BT model are often not transitive, and this is the case particularly when some contrasts $u_i-u_k$ are close to 0. In situations in which the actual data come from a number of individual assessors, as was the case in our musicology experiment, it is a natural idea to try to account in the modeling separately for the two sources that may have created non-transitivity in the combined data: One the one hand, the differences in the assessment profiles of the assessors, and on the other, possible lack of transitivity in the pairwise comparisons coming from each individual assessor. This distinction was made fully explicit in the structure of our BM and BL models of Sections \ref{BM} and \ref{LM}. 

As an alternative to our approach, an anonymous referee suggested a hierarchical two-layer structure based on the BT model. In that suggestion, data coming from an individual assessor would be described by a BT model, but with score parameters $\bm{\mu}_j = (\mu_{j1},...,\mu_{jn})$ specific to each assessor  $j$. On the lower level of model hierarchy, the referee suggested that, for each item $i$, the score parameters $\mu_{ji}$ for different assessors  $j$ would be sampled independently from a Gaussian distribution centered at a common value $\mu_i$.

We developed such a model, which we call HBT (with H for hierarchical), work in progress \citep{hbt}. There, we discuss (i) the suitability of the HBT model for data in the form of repeated pairwise comparisons performed by each assessor, and (ii) the poor performance of HBT compared to our Bayesian Mallows approach when data are such that each assessor only performs a limited number of comparisons without repetitions, so that not all pairs of items are compared by every assessor. Such a small incomplete example, with no intransitivities, is considered in \citet{arsia}, where further differences between the HBT and the Bayesian Mallows model are discussed. 
One important reason for the difference in the case of incomplete and sparse data is that often they do not satisfy the strong connection condition \citep{ford1957solution}. This condition is fulfilled if, for any partition of all items into two non-empty sets, both subsets contain at least one item that was preferred to some item in the other set by at least one assessor, see \citet{yan2016ranking}. If this condition is not satisfied, the maximum likelihood estimator does not exist and the posterior inferences based on the HBT model will be highly sensitive to the specification of the prior and will require corresponding sensitivity analyses.

The BT model was represented and fitted as a log-linear model \citep{dittrich98, dittrich02}. In these works, the authors introduced assessor specific covariates into their framework, and extended it to the case of dependent pair comparisons. Building on \citet{dittrich98}, \citet{francis2010}  further introduced random effects for each assessor in order to account for residual heterogeneity that is not included in individual-specific covariates. However, their method is applied to pair preferences derived from full rankings. As such, the pair preferences are complete, that is, $n(n-1)/2$ pairs are assessed by each assessor, and transitive. 
Their method cannot be used on our data where each assessor provides a limited number of pairwise preferences, typically smaller than the maximum $n(n-1)/2$, and is allowed to contradict herself, thus leading to non-transitive patterns in the data.

An interesting literature that builds on the Thurstone's model is the psychometric one \citep{bockenholt1988logistic, bockenholt2001hierarchical, bockenholt2001individual, bockenholt2006thurstonian}. In these works, the authors develop different generalizations of the Thurstone model, accounting for instance for multidimensional parameters, in case the items are evaluated with respect to multiple aspects, or introducing dependency among the observed pairs, by the inclusion of random effects in the model. 
However inference is performed when the data include repeated comparisons for each assessor, and  all items are compared by each assessor. 

Pair comparison data were also recently handled within the Mallows ranking models by \citet{Lu2015} and \citet{vitelli17}. However, both papers deal only with transitive pairs, explicitly ruling out the non-transitive patterns in the data.
 
\citet{Volkovs2014} propose a score-based method, called Multinomial Preference model (MPM), that generalizes the Plackett Luce model \citep{Luce1959, Plackett1975}.
The main difference between their MPM and our model is in the data generating mechanisms, which \citet{Volkovs2014} assumed to be a multinomial score based process, while our method builds on considering distances between ranking vectors. In addition, their goal is to learn a single consensus ranking of the items, or multiple consensus rankings in case of clustering. Our method instead has the ability to further learn the individual latent rankings for each assessor.

\citet{ding2015learning} proposed a model for noisy pairwise ranking data, based on a mixed membership of Mallows models (M4), which generalizes the mixture model of \citet{Lu2015}. Their proposal is near to ours, in that both postulate the existence of latent linear orderings. However, \citet{ding2015learning} assume a basic separability property, which would be difficult to justify in contexts similar to our data application. Furthermore, they model the presence of non-transitive patterns in the data as arising because each assessor has multiple latent linear orderings, while we propose a mistake model.
Moreover, they consider only the Kendall distance, while our model handles every right-invariant distance.

There is a large body of literature on mixture models for ranking data \citep[e.g][]{MurphyMartin2003, gormley2006analysis, Caron2014, Meila2010, Jacques2014}. Although related to our mixture model extension, all these papers are based on data in the form of rankings, and they do not directly apply, or extend, to non-transitive pairwise comparison data. Apart from this difference, the work of \citet{Jacques2014}, which presents a mixture extension of the model developed in \citet{biernacki2013generative}, has some similarities with ours. These authors assume the existence of a consensus ranking, and of individual rankings, and they model stochastic errors between these permutations of the items, to explain the variability of the individual rankings around the consensus. In this way, the pairwise comparisons are always complete and transitive, in contrast to our setting.


\section{Simulation study}\label{simus}
The aim of the experiments was to validate the method and to evaluate its performance in some test situations. The data were  simulated from the Mallows model with the Bernoulli mistake model, varying parameters $\theta$, $\alpha$, $n$, $N$, and $M_j$, $j=1,...,N$, while always using the footrule distance.
The number of items $n$ was always kept below 50, thus enabling us to use {\color{black}the} exact partition function \citep{vitelli17}. For a detailed description of the data generation, see \ref{suppB}.

Various point estimates can be deduced from the posterior distribution of $\bm{\rho},$ one being the maximum a posteriori (MAP). We prefer the following sequential construction, called the cumulative probability (CP) consensus ordering in \citet{vitelli17}: first we select the item which has the largest marginal posterior probability of being ranked $1^\text{st}$; then, excluding this first choice, we select the item which has the largest marginal posterior probability of being ranked $1^\text{st}$ or  $2^\text{nd}$ among the remaining ones, and so on. 

\begin{figure}[!ht]
\centering
\centerline{\includegraphics[width=0.9\textwidth]{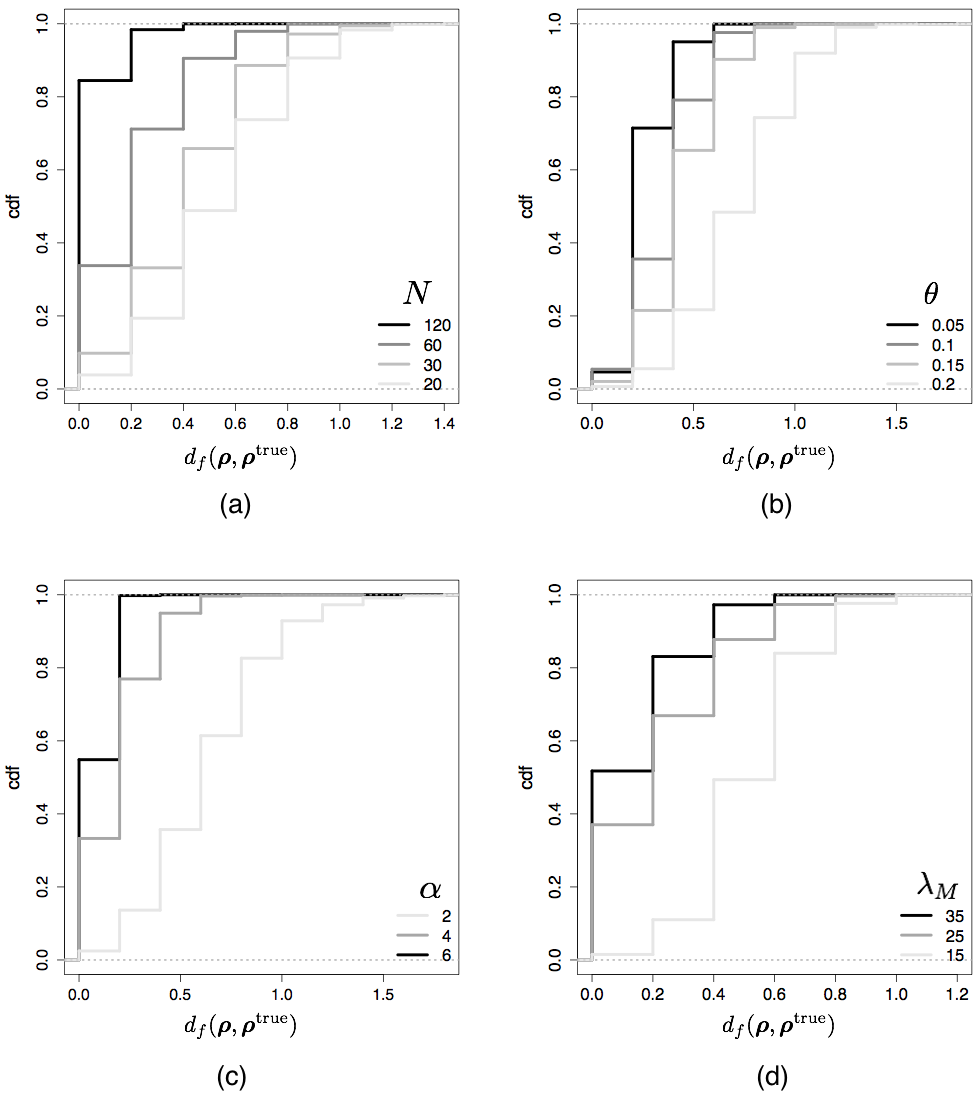}}
\caption{Results of the simulated data. Posterior CDFs of $d_f(\bm{\rho}, \bm{\rho}^{\text{true}})$  as a  function of $N$ for $\alpha=3$, $\lambda_M=25$, $\theta=0.1$ (a); as a function of $\theta$ for $\alpha=3$, $N=40$, $\lambda_M=25$  (b); as a function of $\alpha$ for $\theta=0.1$, $N=40$, $\lambda_M=25$  (c); as a function of $\lambda_M$ for $\alpha=3$, $N=40$, $\theta=0.1$  (d).}
\label{cdf1}
\end{figure}

In order to assess the performance of our methods, in Figure \ref{cdf1}  we plot the posterior distribution of the normalized footrule distance between the estimated consensus $\bm{\rho}$ and the true consensus, $d_f(\bm{\rho}, \bm{\rho}^{\text{true}})=\frac{1}{n}\sum_{i=1}^n|\rho_i-\rho_{i}^{\text{true}}|$, for varying  parameters $\alpha$, $\theta$, $\lambda_M$ (the average number of pairs given to each assessor) and $N$, while keeping fixed $n=10$.  

As  expected, the performance of the method improves as the number of assessors $N$ increases (Figure  \ref{cdf1}a),  as the probability of making mistakes $\theta$ decreases (Figure  \ref{cdf1}b), as the dispersion of the individual latent rankings $\bm{R}_{j}^{\text{true}}$ around $\bm{\rho}^{\text{true}}$ decreases, that is when $\alpha$ increases (Figure  \ref{cdf1}c),  and when the average number of pairwise comparisons becomes larger (Figure  \ref{cdf1}d). Interestingly, in the last case, the method performs generally well also when the average number of pairs is $\lambda_M=15$, being only 1/3 of the maximal number of pairs possible. 

\begin{figure}[h!]
\centering
\centerline{\includegraphics[width=0.9\textwidth]{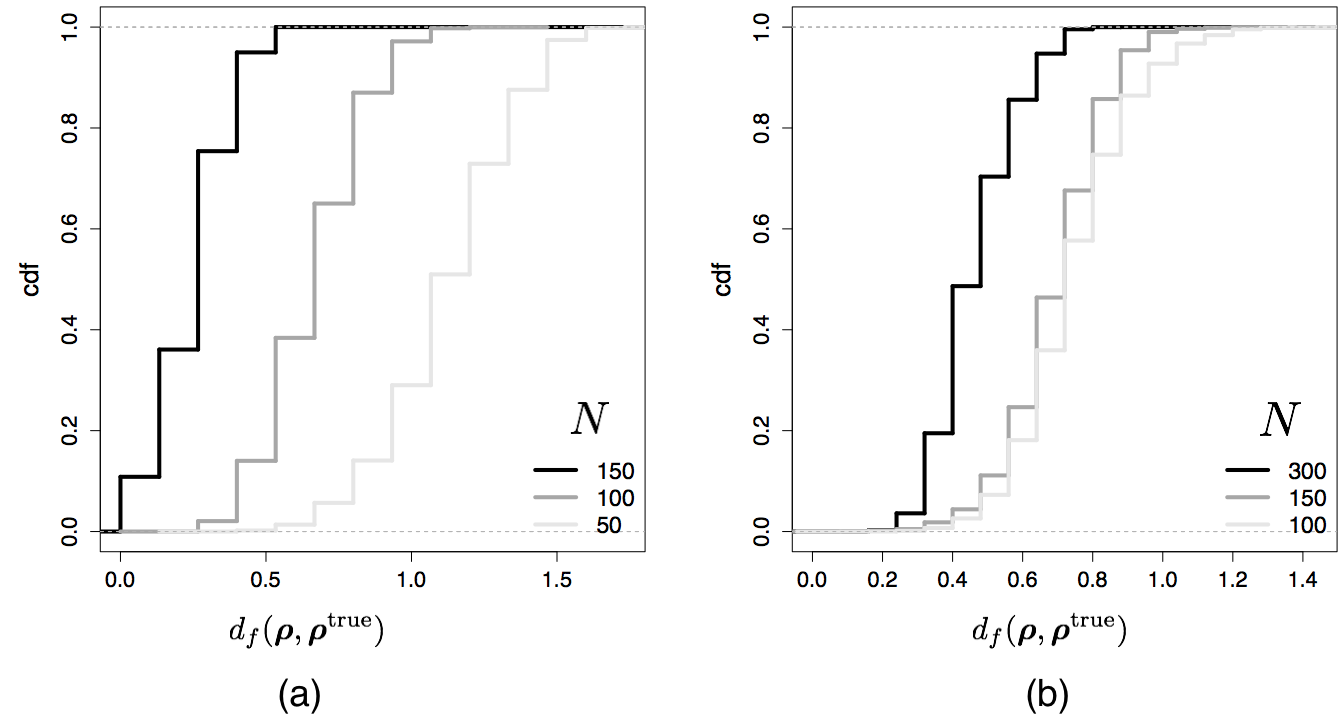}}
\caption{Results of the simulated data.  Posterior CDFs of $d_f(\bm{\rho}, \bm{\rho}^{\text{true}})$    as a  function of $N$, for $\theta=0.1$, $\alpha=3.5$, $\lambda_M=25$, $n=15$ (a), {and} for $\theta=0.1$, $\alpha=4.5$, $\lambda_M=50$, $n=25$ (b). }
\label{cdd}
\end{figure}

In Figure \ref{cdd} we plot the posterior distribution of  $d_f(\bm{\rho}, \bm{\rho}^{\text{true}})$
corresponding to simulation experiments with $n\in \{15,25\}$, when increasing the number of assessors $N$.
Note that the number of pairs assessed by each assessor in the case $n=25$ is around 50, which is 1/6 of all the possible pairs.  

Next, we studied the performance of the method in terms of the precision of the individual ranking estimation.
We quantified the results by the probability of getting at least 3 items right, among the top-$5$, defined as follows.
For each assessor $j=1,...,N$, we found the triplet of  items $D_3^j=\{O_{i_1}, O_{i_2}, O_{i_3}\}$ that had maximum posterior probability of being ranked jointly among the top$-3$ items, i.e. the triplet that maximized $\sum_{\sigma\in\mathcal{P}_{3}}\mathbb{P}(\{R_{j{\color{black}i_1}}, R_{j{\color{black}i_2}}, R_{j{\color{black}i_3}}\} = \sigma \,|\, \text{data}),$ where $\sigma$ denotes a permutation of the set ${\{1,2,3\}}$.
This posterior quantity was estimated along the MCMC trajectory.
We defined $H_5^j$ to be the set of 5 highest ranked items in $\bm{R}_{j}^{\text{true}}$, for each assessor $j$. We then checked whether $D_3^j\subset H^j_5$ (that is, if the top-3 estimated items were all among the top-5 of each assessor). The percentages of assessors for which this is true are reported in Table \ref{top3}. 
We notice that the results are overall very good: in the cases where $n$ is set to 10 (first 4 sub-tables from the left in Table \ref{top3}), we consistently learn 3 out of the top$-5$ items in more than 70\% of the assessors (with a peak of 100\%). Also in the more difficult cases of $n=15$ and $n=25$ (first 2 sub-tables from the right in Table \ref{top3}) the results are very good, especially considering that this percentage does not include the cases where only 2 (or 1) items where correctly estimated in the top positions.

\begin{table}[h!]
\caption{Results of the simulated data. Percentage of assessors for which the estimated top-$3$ items belong to the true top-$5$. Data correspond to simulations with the same parameter settings as the results shown in Figures \ref{cdf1} and \ref{cdd}: from left to right, same parameters as in Figure \ref{cdf1}a, Figure \ref{cdf1}b, Figure \ref{cdf1}c, Figure \ref{cdf1}d, Figure \ref{cdd}a and Figure \ref{cdd}b.}
\centering
        \begin{tabular}{|c|c|c|c|c|c|c|c|c|c|c|c|c|c|c|c|c|}
 \cline{1-2}\cline{4-5}\cline{7-8}\cline{10-11}\cline{13-14}\cline{16-17}
            $N$ & \% & & $\theta$&\%& & $\alpha$ & \% &&$\lambda_{M}$&\% &&$N$&\% &&$N$&\%  \\ 
 \cline{1-2}\cline{4-5}\cline{7-8}\cline{10-11}\cline{13-14}\cline{16-17}
         20 & 88 &&0.05 & 92.5&&2 & 82.5&&15 & 85  &&50 & 65 && 100 & 44 \\  
 \cline{1-2}\cline{4-5}\cline{7-8}\cline{10-11}\cline{13-14}\cline{16-17}
	30 & 83 &&0.1 & 87.5 &&4 & 95&&25 & 97.5  &&100 & 58 &&150 & 46 \\  
 \cline{1-2}\cline{4-5}\cline{7-8}\cline{10-11}\cline{13-14}\cline{16-17}
 	60 &83&&0.15 & 75 &&6 &92.5 && 35 & 100 &&150 & 60 &&300 & 45 \\ 
 \cline{1-2}\cline{4-5}\cline{7-8}\cline{10-11}\cline{13-14}\cline{16-17}
	120&75&&0.2& 72.5 & \multicolumn{12}{c}{ } \\  
 \cline{1-2}\cline{4-5}
 \end{tabular}            

\label{top3}
\end{table}

\begin{figure}[h!]
\centering\includegraphics[width=\textwidth]{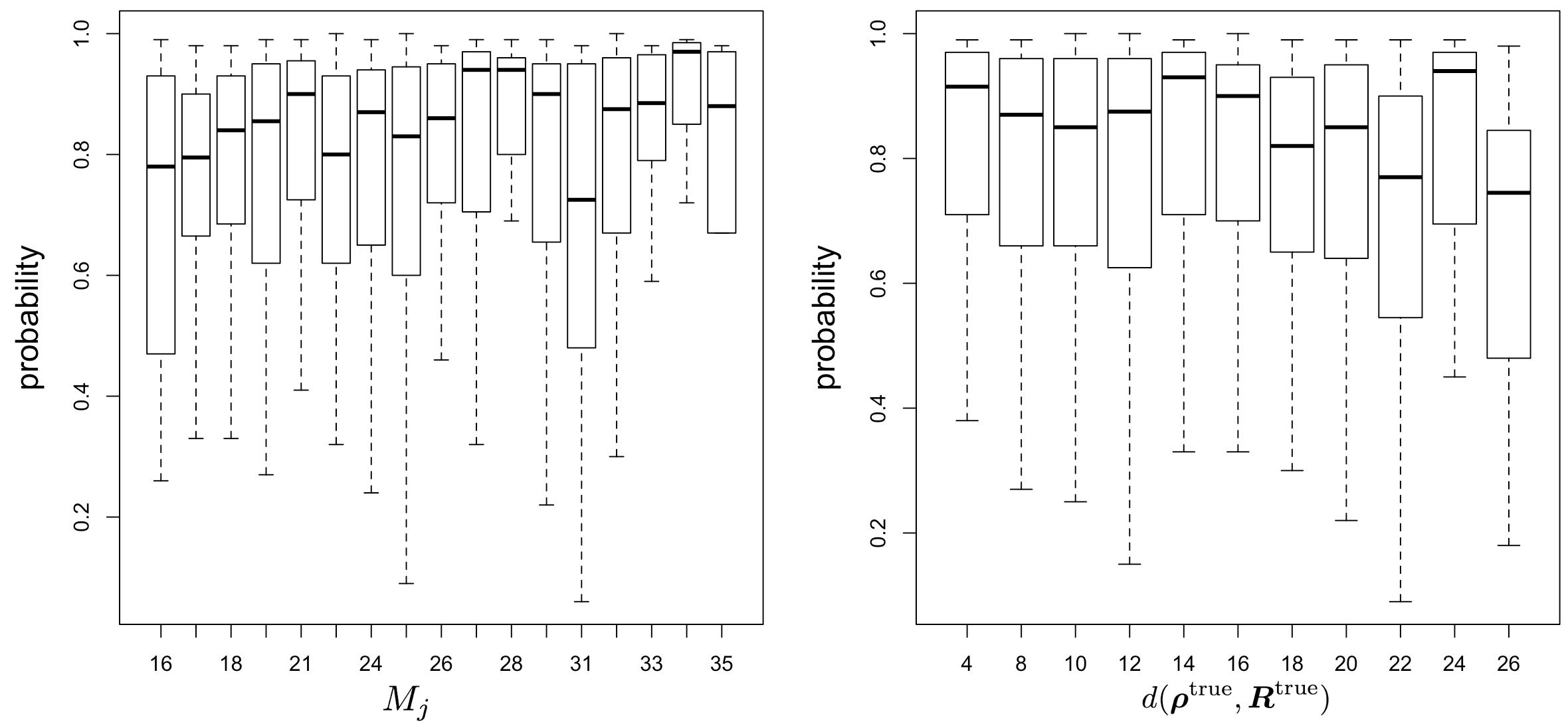}
\caption{{\color{black}Results of the simulated data.} Posterior  probabilities of correctly predicting the preference order of all pairs not assessed by the assessors, (left) stratified according to the number of pairs each assessor assessed in the data, and (right)  stratified according to $d(\bm{\rho}^\text{true},\bm{R}_{j}^{\text{true}})$.
}
\label{rtildeyo2}
 \end{figure}

We then chose randomly one of the simulated data cases and computed the posterior  probabilities of correctly predicting the preference order of all pairs not assessed by the assessors, i.e.                                    
$\mathbb{P}\Big[g({\mathcal{B}}_{j,\text{new}},\bm{R}_j)=g({\mathcal{B}}_{j,\text{new}},\bm{R}_{j}^{\text{true}})\,\Big|\,\text{data}\Big].
$
Figure \ref{rtildeyo2} shows the boxplots for these predictive probabilities, (left) stratified according to the number of pairs each assessor assessed in the data, and (right)  stratified according to the footrule distance between the individual ranking $\bm{R}_{j}^{\text{true}}$ and the consensus $\bm{\rho}^\text{true}$, $d(\bm{\rho}^\text{true},\bm{R}_{j}^{\text{true}})=\sum_{i=1}^n|\rho_{i}^{\text{true}}-R_{ji}^{\text{true}}|$.

In the case considered, the model had a very good predictive power, especially considering that the simulated data had many mistakes (around 10\%). We also notice a slight increase of the predictive probabilities as $M_j$ increases (left panel) and as $d(\bm{\rho}^\text{true},\bm{R}_{j}^{\text{true}})$ decreases (right panel). 
These results are not surprising: it is easier to predict correct orderings of new pairs when (i) the assessor assesses more pairs, and (ii) the assessor's own ranking resembles more the shared consensus.   

In \ref{suppC} we report an analysis of data generated by the logistic model LM. 
 The results were very similar to those obtained above. In fact, the posterior distribution of $\beta_1$ was highly concentrated around 0, which is when LM collapses to BM.

\section{Human causation in sounds}\label{res_soni}
We analyzed the data using the mixture model explained in Section \ref{mistura} with  footrule distance. With $n=12$ sounds we can use the exact expression of the partition function \citep{vitelli17}.
In the Dirichlet prior for $\eta$, we set $\chi = 20$, which favors high-entropy distributions, thus reflecting our inability to express precise prior knowledge.
In the Beta prior for $\theta$, we set the hyperparameters at $\kappa_1=\kappa_2=1$, that is, the uniform distribution on the interval $[0,0.5)$, and the hyperparameters of the prior for $\alpha$ at $\gamma=1$ and $\lambda=1/10$, as discussed in \citet{vitelli17}.  We run the MCMC sampler for $10^6$ iterations, after a burn-in of $2\cdot10^5$.
Separate analyses were performed for $G \in \{1,\dots,7\}$. 

\begin{figure}[h!]
\includegraphics[width=0.45\linewidth]{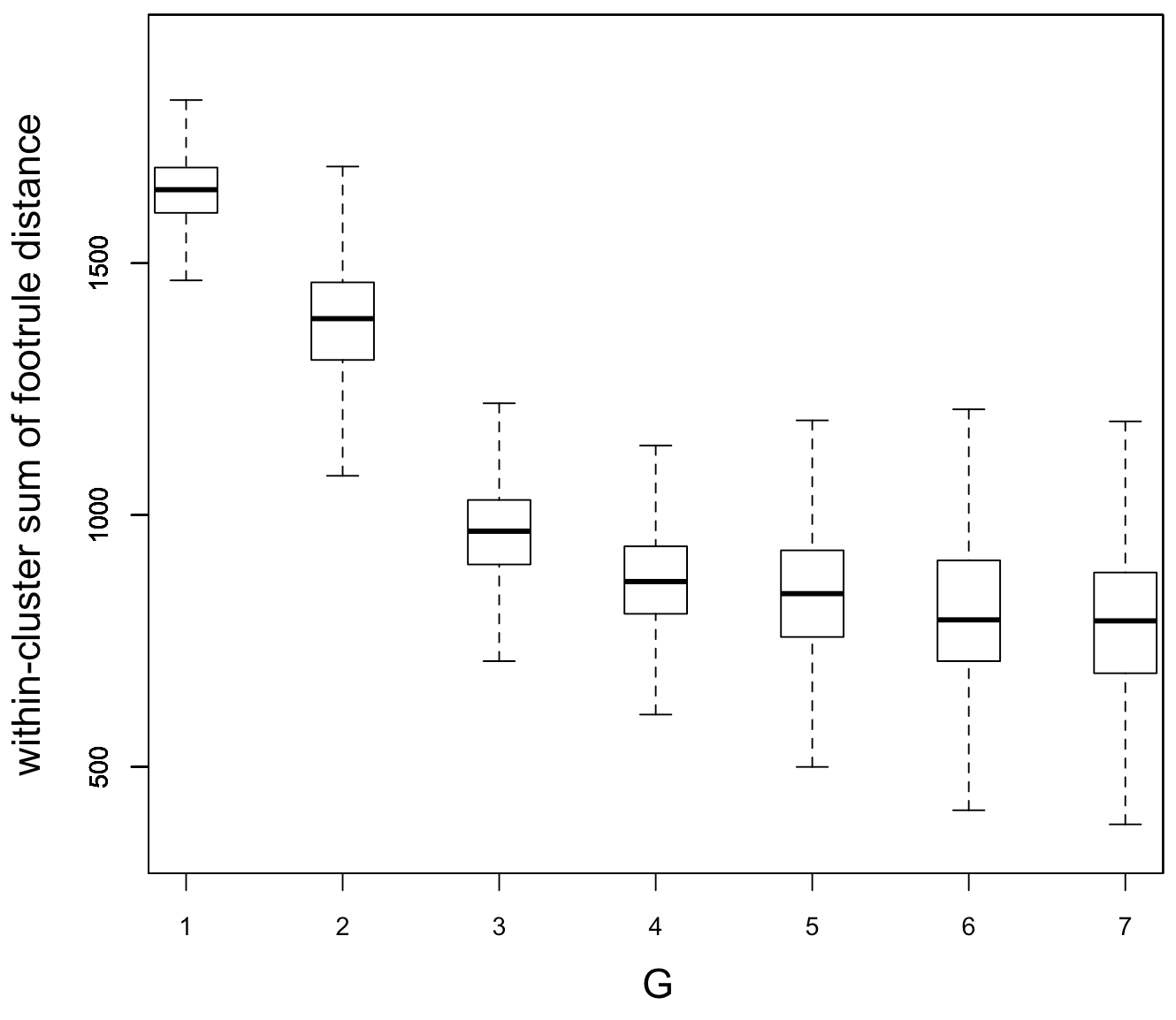} 
\includegraphics[width=0.45\linewidth]{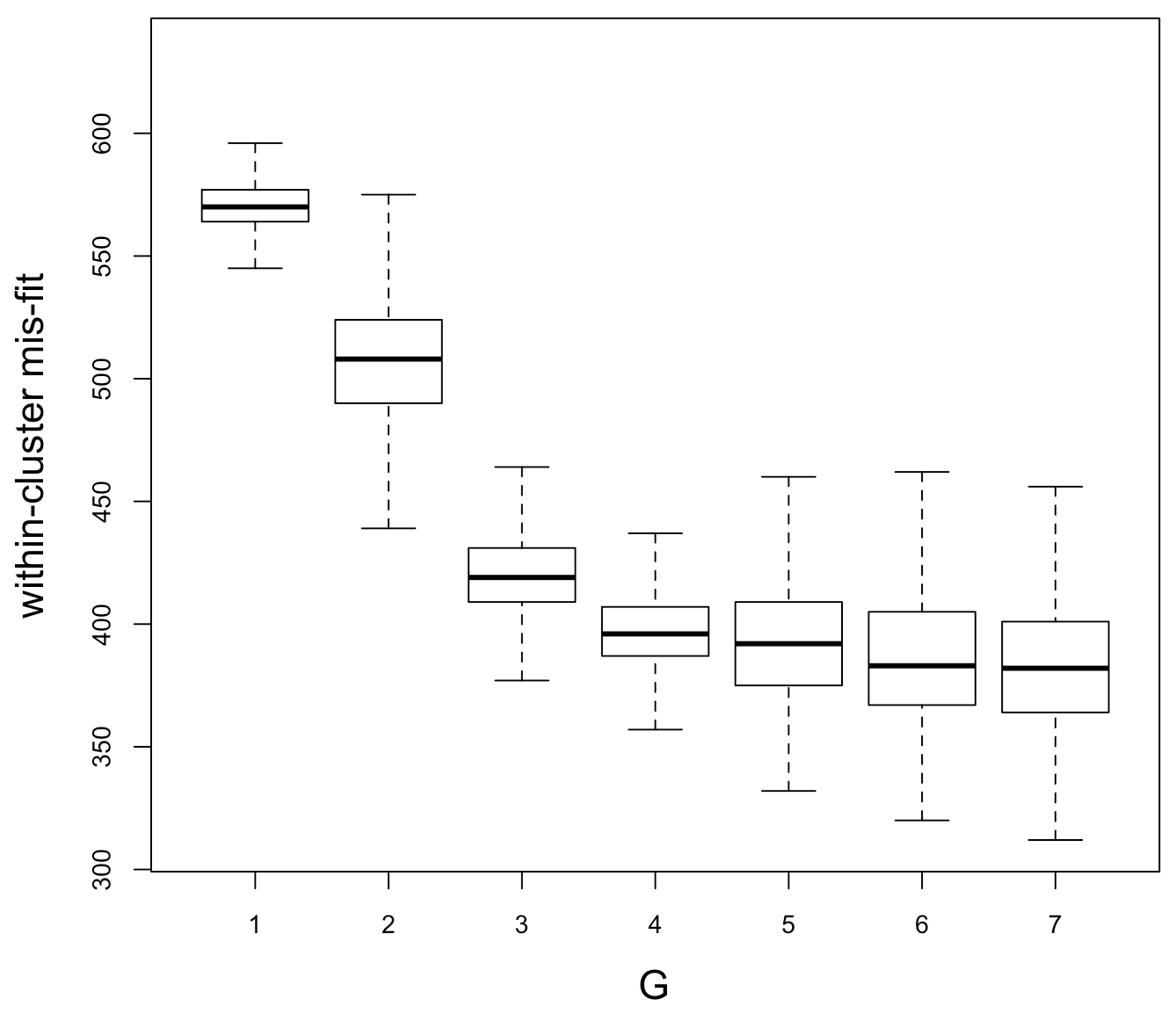} 
\caption{\emph{Acousmatic} data. Boxplots of the within-cluster sum of footrule distances between the individual rankings and the consensus ranking of that cluster (left), and of the within-cluster indicator of mis-fit to the data (right), for different choices of $G$.}\label{clust}
\end{figure}

In order to choose an appropriate number of clusters,  we plot in  Figure \ref{clust} two quantities:  on the left,
the within-cluster sum of footrule distances between the individual rankings and the consensus ranking of that cluster,
$\sum_{g =1}^G\sum_{j:z_j = g}d_f(\bm{R}_j,\bm{\rho}_g)$; 
on the right, the within-cluster indicator of mis-fit to the data, $\sum_{g =1}^G\sum_{j:z_j = g}\sum_{m=1}^{M_j} \text{g}(\mathcal{B}_{jm},\bm{\rho}_g)$. Both these measures are defined in \citet{vitelli17}, and tested as good measures to select $G$. 

More traditional information criteria, such as the deviance information criterion \citep{Spiegelhalter2002}, were considered, however their performance was quite unstable, possibly attributable to the sparsity of the data.\\
Inference on the number of clusters could have been alternatively performed via a reversible jump MCMC.

There appears to be  an elbow in the figures at $G=3$, to guide us in the choice of the number of clusters. We decided on $G=3$, also motivated by the relatively small sample size of the experiment ($N=46$).

Table \ref{mixtTable} shows the results for $G = 3$: the maximum a posteriori (MAP) estimates for $\eta$ and $\alpha$, together with their 95\% highest posterior density (HPD) intervals, are shown at the top of the table. The table also shows the estimated cluster-specific consensus lists of sounds, estimated by the CP procedure.
We observe the differences in the three consensus lists. S1, the stimulus with the most dynamic spatial motion, is on top in cluster 3, but at the bottom in cluster 1; S8, the test stimulus that has maximum spatial details but no volume nor pitch change, is on top in cluster 1, but second to the last in clusters 2 and 3. Finally, S5, the stimulus that contains the least movement variation but has pitch and volume suppressed, is ranked third and first in clusters 1 and 2, but towards the bottom of the list in cluster 3.

\begin{table}[h!]
\begin{tiny}
\centering
\caption{\texttt{Acousmatic} data. Sounds are ordered according to the CP consensus ordering, obtained from the posterior distribution of $\bm{\rho}_g$, $g = 1,2,3$.}\centering
\label{mixtTable}
\begin{tabular}{l|c|l|l|l|l|l|l|l|l|l|l|l|l|}
\cline{2-14}
                                                    &                           \textbf{Rank}                     & \textbf{1}          & \textbf{2}           & \textbf{3}          & \textbf{4}          & \textbf{5}          & \textbf{6}          & \textbf{7}          & \textbf{8}           & \textbf{9}           & \textbf{10}         & \textbf{11}         & \textbf{12}         \\ \hline
\multicolumn{1}{|l|}{\multirow{2}{*}{\textbf{G1}}} & \begin{tabular}[c]{@{}l@{}}$\alpha_1=$ 2.66\\ (1.14,4.96)\end{tabular} & \multirow{2}{*}{S8} & \multirow{2}{*}{S10} & \multirow{2}{*}{S5} & \multirow{2}{*}{S9} & \multirow{2}{*}{S6} & \multirow{2}{*}{S4} & \multirow{2}{*}{S7} & \multirow{2}{*}{S11} & \multirow{2}{*}{S12} & \multirow{2}{*}{S2} & \multirow{2}{*}{S3} & \multirow{2}{*}{S1} \\ \cline{2-2}
\multicolumn{1}{|l|}{}                             & \begin{tabular}[c]{@{}l@{}}$\eta_1$=0.31\\ (0.21,0.41) \end{tabular}  &                     &                      &                     &                     &                     &                     &                     &                      &                      &                     &                     &                     \\ \hline\hline
\multicolumn{1}{|l|}{\multirow{2}{*}{\textbf{G2}}} & \begin{tabular}[c]{@{}l@{}}$\alpha_2=$ 5.16\\ (3.15,9.29)\end{tabular} & \multirow{2}{*}{S5}   & \multirow{2}{*}{S4}    & \multirow{2}{*}{S12}   & \multirow{2}{*}{S2}   & \multirow{2}{*}{S11}   & \multirow{2}{*}{S3}   & \multirow{2}{*}{S6}   & \multirow{2}{*}{S1}    & \multirow{2}{*}{S7}    & \multirow{2}{*}{S9}   & \multirow{2}{*}{S8}   & \multirow{2}{*}{S10}   \\ \cline{2-2}
\multicolumn{1}{|l|}{}                             & \begin{tabular}[c]{@{}l@{}}$\eta_2$=0.33\\ (0.22,0.43)\end{tabular}  &                     &                      &                     &                     &                     &                     &                     &                      &                      &                     &                     &                     \\ \hline\hline
\multicolumn{1}{|l|}{\multirow{2}{*}{\textbf{G3}}} & \begin{tabular}[c]{@{}l@{}}$\alpha_3=$5.32 \\(3.61,7.66) \end{tabular} & \multirow{2}{*}{S1}   & \multirow{2}{*}{S7}    & \multirow{2}{*}{S11}   & \multirow{2}{*}{S2}   & \multirow{2}{*}{S4}   & \multirow{2}{*}{S12}   & \multirow{2}{*}{S6}   & \multirow{2}{*}{S3}    & \multirow{2}{*}{S5}    & \multirow{2}{*}{S9}   & \multirow{2}{*}{S8}   & \multirow{2}{*}{S10}   \\ \cline{2-2}
\multicolumn{1}{|l|}{}                             & \begin{tabular}[c]{@{}l@{}}$\eta_3$=0.37\\ (0.27,0.48)\end{tabular}  &                     &                      &                     &                     &                     &                     &                     &                      &                      &                     &                     &                     \\ \hline
\end{tabular}
\end{tiny}
\end{table}

Listeners in cluster 1 found variation in volume or pitch as a negative or distracting feature. They rated S8 at the top, a test stimulus that has maximum spatial details but no volume nor pitch change. Also, S10, S5 and S9, which were ranked next, lack volume and pitch details. The bottom 4 stimuli contain maximum pitch and volume variation. Among them was S3 (mono sound, no space at all), forming a strong contrast to the top ranked S8  (maximum spatial movement). Evidently, space was important for these listeners, while pitch and volume variation was a negative or distracting feature. 

In cluster 2 listeners did not like fast movements as a sign of human feature, but they did like correlated pitch and volume (the top 4 sounds feature a low amount of spatial variation, but also correlated pitch and volume, while the bottom 3 sounds are the same as the top 3 but lack correlated pitch and volume variation). Listeners in this cluster prioritized pitch and volume variations above spatial variation, and preferred low spatial variation (slower, or more relaxed movements). 

Cluster 3 consists of subjects who, in their evaluation of the test stimuli, appear to include all spatial cues that adhere to our everyday perception of spatial motion. The stimuli with most dynamic spatial motion, enhanced by spatially correlated pitch and volume variations, are in the top-3, while stimuli with the least of these features are in the bottom-3. These listeners prioritize high levels of spatial detail above all other features, and their perception of these details are enhanced by correlated pitch and volume variations. This is indicated in (i) S1 being at the top; (ii) S7, which is the same as S1 but lacks pitch variation, being second; (iii) S11, which is the same as S1 but played 30\% slower, being third (i.e. space, volume and pitch variations are just a bit slower); (iv) S8, S9, S10 are in the bottom, and all lack pitch, volume variation, and  spatial movement details.

\begin{figure}[ht]\includegraphics[width=\linewidth]{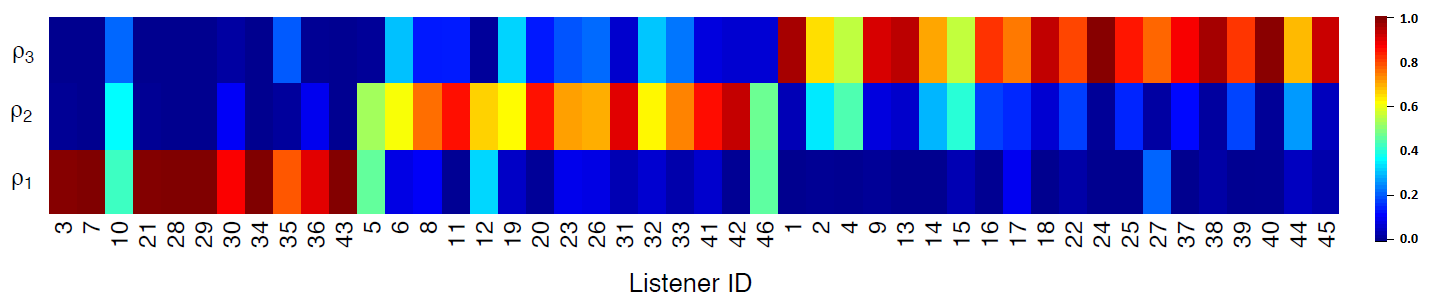} 
\caption{{\color{black}\emph{Acousmatic} data.} Heatplot, for all the listeners (on the x-axis), of the posterior probabilities of being assigned to each of the three clusters (on the y-axis).}\label{stab}
\end{figure}

We investigate the stability of the clustering in Figure \ref{stab}, that shows the heatplot of the  posterior probabilities, for all the listeners (shown on the x-axis), for being assigned to each of the clusters identified in Table  \ref{mixtTable}. Most of the probabilities are concentrated on some particular  value of $c$ among the three possibilities, indicating a reasonably precise behavior in the cluster assignments. 

\begin{figure}[h!]
\includegraphics[width=\linewidth]{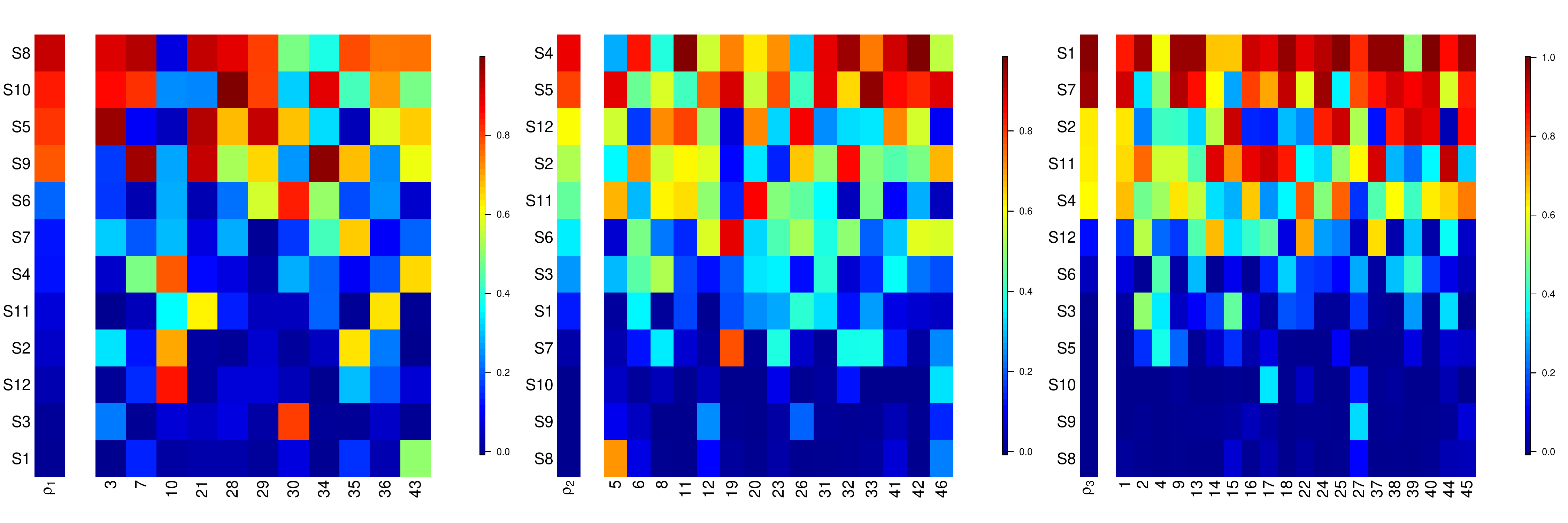} 
\caption{\emph{Acousmatic} data. Heatplot of the marginal posterior probabilities for all the stimuli (y-axis) of being ranked among the top-4 for cluster 1 (left), 2 (center) and 3 (right). }\label{stab2}
\end{figure}

We then computed, fixing these cluster assignments, the marginal posterior probability that each sound is among the top-4 in $\bm{\rho}_{1:G}$  and in $\bm{R}_j$, $j=1,...,46$, respectively.
The results  are shown in Figure \ref{stab2}. 
Each heatplot refers to a cluster (G1 (left), G2 (center) and G3 (right)) and represents the marginal posterior probabilities for each sound (y-axis) being ranked among the top-4 in the consensus of that cluster (first column), and in the individual rankings of listeners in that cluster (remaining columns, assessors on the x-axis). As Figure \ref{stab2} shows, there is considerable variation in the estimated rankings of the sounds between individual listeners even when they are included in the same cluster. 
For example, looking at Figure \ref{stab2} left, we see that S8, S10, and S5  have high  ($>0.8$) posterior probability of being ranked among the top-4 stimuli in the consensus ranking (column 1). However, looking at the estimates for the listeners in cluster 1, we see that the variation is very high: For example, listener 30 (column with label 30) has a very high  posterior probability of ranking S3 and S6 among the top-4 stimuli.
This aspect is important for what concerns individual estimates.

\begin{figure}[h!]
\includegraphics[width=\linewidth]{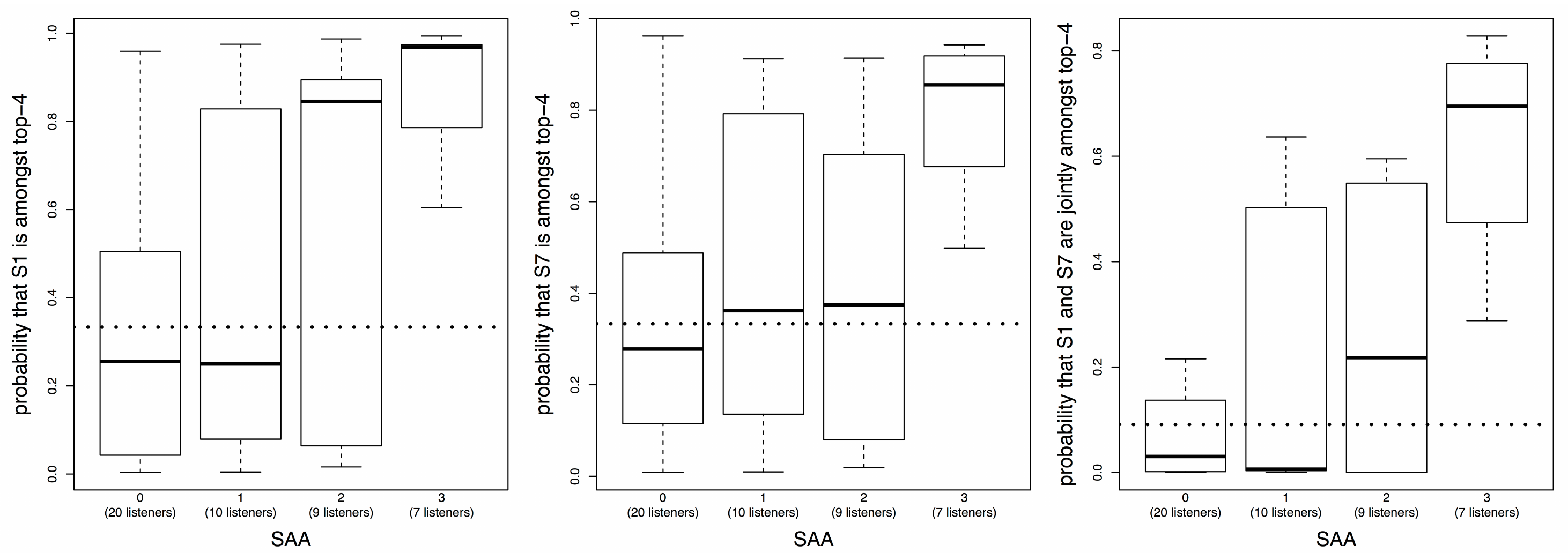} 
\caption{\emph{Acousmatic} data. Boxplot of the posterior probabilities for sounds S1 (left), S7 (middle), S1 and S7 jointly (right), of being ranked among the top-4 in the individual ranking $\bm{R}_j$, stratified by the SAA index. The horizontal dotted line is the threshold in the case of random assignment. Scale of SAA: from 0 to 3, the largest, the more aware of spatial dimension of sounds.}\label{SAA17}
\end{figure}

Here we consider the relationship between the probability of placing some given stimuli in the top (bottom) ranks and the musical sophistication index (MSI), or the spatial audio awareness index (SAA). 
Figure \ref{SAA17} shows the relationship between  listeners' SAA and the probability of  sounds S1 and S7  being ranked in the top-4 (both marginally and jointly). Recall that S1 was the original sound, while S7 was identical to  S1, but without pitch variation.
The plot suggests that spatial listening is a skill that is enhanced through training. 

 \begin{figure}[h!]
\includegraphics[width=\linewidth]{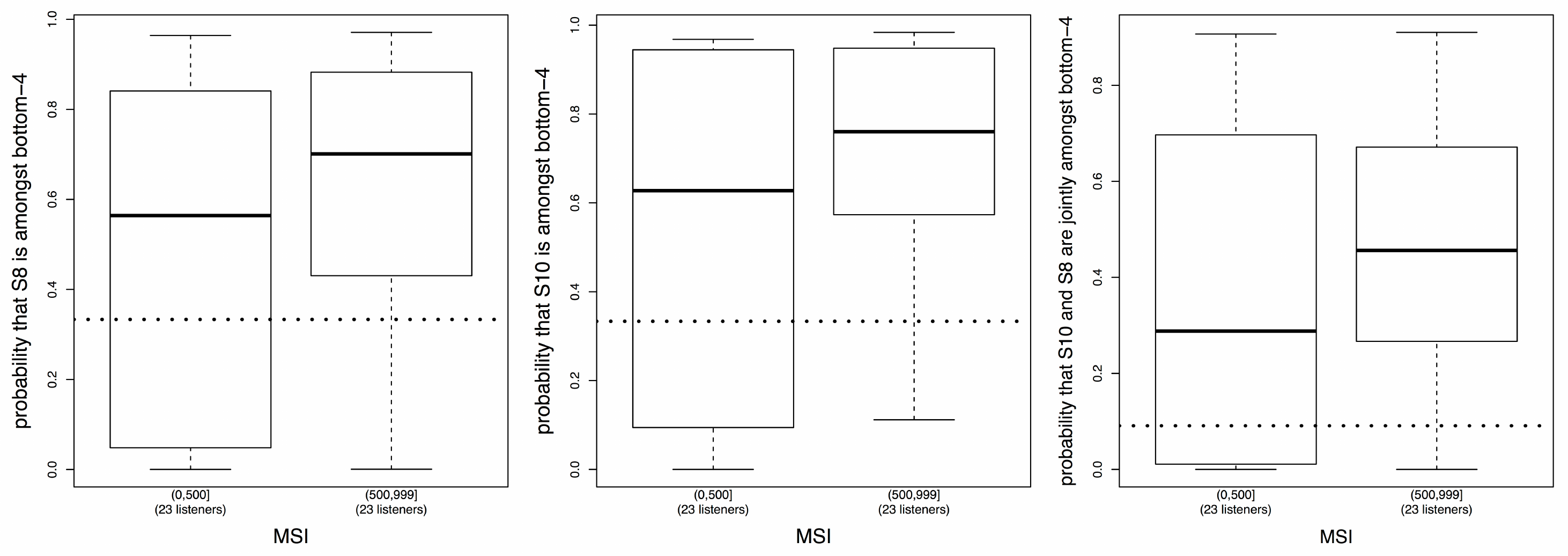} 
\caption{\emph{Acousmatic} data. Boxplot of the posterior probabilities for sounds S8 (left), S10 (middle), S8 and S10 jointly (right), of being ranked among the bottom-4 in the individual ranking $\bm{R}_j$, stratified by the MSI index. The horizontal dotted line is the threshold in the case of random assignment.}\label{MSI17}
\end{figure}

Figure \ref{MSI17} shows the relationship between  listeners' MSI and the probability of  sounds S8 and S10  being ranked among the bottom-4 (both marginally and jointly). 
Respondents with a score greater than 500 were classified as  musically more sophisticated, and those with a score less than 500 as less sophisticated, as suggested in \texttt{http://marcs-survey.uws.edu.au/OMSI/omsi.php}. Both S8 and S10 suppress pitch and volume variations, which are expected to enhance the implication of human causation. These two stimuli are more likely to be ranked in the last 4 positions by listeners with high MSI.
Interestingly, this suggests that musically sophisticated listeners find pitch and volume variations to be qualities for a stimulus to sound human.

\section{Conclusions and discussion}\label{concl}
The main contribution of this paper is to introduce a new Bayesian method for non-transitive pairwise preference data. The principal advantage of the Bayesian approach comes from its ability to combine different types of uncertainty in the reported data, coming from different sources, and from being able to convert such data {into} the form of meaningful probabilistic inferences. Our method provides the posterior distribution of the consensus ranking, based on pairwise assessment data from a pool of  assessors who may have individually violated logical transitivity in their reporting. The method is also able to produce the posterior distributions of the latent  individual rankings of the assessors. Such rankings can  be used in the construction of personalized recommendations, or in studying how individual preferences change with assessor related covariates.
We also developed a mixture model generalization of the main model, able to handle heterogeneity in pairwise and non-transitive preference data. 
The model was then used to investigate how individual listeners perceive human spatial causation in \emph{acousmatic} sounds. The data came from a difficult experiment, that involved human perceptions. 
For this reason, pair comparison of sounds was the only feasible design. The data were noisy, and in particular often logically non-transitive at the individual level. We used our approach to estimate individual rankings, and sub-groups of assessors. The results revealed how differently people listen to and interpret abstract sounds. We related individual musicological scores to individual rankings, leading to an interesting correspondence between spatial sound feelings and sound expertise.

Sometimes pairwise comparison data contain draws, or ties. 
A tie occurs when a pairwise comparison between two items does not result in a defined preference of an item towards the other.  This situation has been much considered in the literature on pairwise comparisons \citep[e.g.][]{rao1967ties, davidson70}. 
Our method does not model probabilistically the presence of ties, but it is possible to handle them directly in the MCMC procedure: apply the proposed model, and simply break each tie by tossing a symmetric coin inside the MCMC.  

Another extension of the model would be to allow for the possibility of including covariates of subjects and/or items in the analysis. For instance, the probability of making a mistake could depend on some characteristics of the items, so that, the more similar two items are in terms of such characteristics, the more likely it is to make a mistake in reporting the pairwise preference. In our application relevant covariates could be the variation in pitch and volume, or the overall motion speed that characterizes each sound.

The time complexity of our algorithm is linear in terms of the number of assessors $N$. The increase of the number of items $n$ does not affect computing time of a single MCMC step. However, the larger $n$ is, the longer the chain must be in order to reach convergence.

\section*{Acknowledgments}
{MC thanks Sonia Petrone and Isadora Antoniano-Villalobos  for useful discussions. MC was partially funded by Cariplo during the project. The authors thank the Editor and the anonymous reviewers for their useful comments and suggestions which helped improving the paper. We thank \O ystein S\o rensen for his contributions to Bayesian Mallows modeling. }

\begin{supplement}[id=suppA]
  \sname{Supplement  A}
  \stitle{Adaptations of the algorithm}
\slink[doi]{10.1214/00-AOASXXXXSUPP}
\end{supplement}

\begin{supplement}[id=suppB]
  \sname{Supplement  B}
  \stitle{Some remarks on the simulated data}
  \slink[doi]{10.1214/00-AOASXXXXSUPP}
 \end{supplement}

\begin{supplement}[id=suppC]
  \sname{Supplement  C}
  \stitle{Simulations with the Logistic DGP}
  \slink[doi]{10.1214/00-AOASXXXXSUPP}
\end{supplement}

\begin{supplement}[id=suppD]
  \sname{Supplement  D}
  \stitle{MCMC diagnostics}
  \slink[doi]{10.1214/00-AOASXXXXSUPP}
\end{supplement}

\bibliography{referencesTesi}
\bibliographystyle{imsart-nameyear}

\end{document}